\documentclass[aps,twocolumn,groupedaddress,showpacs]{revtex4}
\usepackage{graphics}                                                           
\usepackage{color}                                                              
\usepackage{ifthen}                                                             
\usepackage{amssymb}
\usepackage{mathrsfs}
\usepackage{float}
\usepackage{bbold}
\usepackage{tabularx}
\usepackage{mathtools}
\usepackage{cellspace,tabularx}
\usepackage{array}
\usepackage{multirow}

\def\nn{\nonumber}
\def\ds{{\rm d}s}
\def\da{{\rm d}a}

\def\dV{{\rm d}V}

\def\W{\mathcal W}
\def\R2{{\mathcal R}^2}
\def\SS{{\mathcal S}^2}

\def\be{\begin{equation}}
\def\ee{\end{equation}}
\def\barr{\begin{array}{lr}}
\def\earr{\end{array}}
\def\bea{\begin{eqnarray}}
\def\eea{\end{eqnarray}}
\def\la{\langle}
\def\ra{\rangle}
\def\vx{{\mathbf x}}

\def\nn{\nonumber}

\def\a{\alpha}

\def\d{\delta}

\def\g{\gamma}

\def\l{\lambda}

\def\s{\sigma}
\def\t{\tau}

\def\x{\xi}

\def\D{\Delta}

\def\ha{\frac{1}{2}}
\definecolor{maroon}{rgb}{0.5, 0.0, 0.0}	
\definecolor{arsenic}{rgb}{0.23, 0.27, 0.29}

\begin{document}
\title{The geometrical meaning of statistical isotropy of smooth random fields in two dimensions}

\author{Pravabati Chingangbam$^{1}$}                                          
\email{prava@iiap.res.in}

\author{Priya Goyal$^{1,2}$}                                          

\author{K. P. Yogendran$^{3}$}                                                

\author{Stephen Appleby$^{4,5}$}                                                

\affiliation{$^1$ Indian Institute of Astrophysics, Koramangala II Block,  
  Bangalore  560 034, India\\
        $^2$ Department of Physics, Pondicherry University, R.V. Nagar, Kalapet, 605014, Puducherry, India\\
  $^3$ Department of Physics, Indian Institute of Science Education and Research, Sector 81, Mohali,  India\\
$^4$ Asia Pacific Center for Theoretical Physics, Pohang, 37673, Korea \\
$^{5}$ Department of Physics, POSTECH, Pohang 37673, Korea}

\begin{abstract}
We revisit the geometrical meaning of statistical isotropy that is manifest in excursion sets of smooth random fields in two dimensions.  
Using the contour Minkowski tensor, $\W_1$, as our basic tool we first examine geometrical properties of single structures. For simple closed curves in two dimensions we show that $\W_1$ is proportional to the identity matrix if the curve has $m$-fold symmetry, with $m\ge 3$. Then we elaborate on how $\W_1$ maps any arbitrary shaped simple closed curve to an ellipse that is unique up to translations of its centroid. We also carry out a comparison of the shape parameters, $\alpha$ and $\beta$,  defined using $\W_1$, with the filamentarity parameter defined using two scalar Minkowski functionals - area and contour length. We show that they contain complementary shape information, with $\W_1$ containing additional information of orientation of structures.
Next, we apply our method to boundaries of excursion sets of random fields and examine what statistical isotropy means for the geometry of the excursion sets.  Focusing  on Gaussian isotropic fields, and using a semi-numerical approach we quantify the effect of finite sampling of the field on the geometry of the excursion sets. In  doing so we obtain an analytic expression for $\alpha$ which takes into  account the effect of finite sampling. Finally we derive an analytic expression for the ensemble expectation of $\W_1$ for Gaussian anisotropic random fields. Our results provide insights that are useful for designing tests of statistical isotropy using cosmological data. 
 
\end{abstract}

\maketitle
\section{Introduction}
\label{sec:intro}

The symmetry properties and statistical nature of cosmological fields 
are central to our understanding of the universe. The $\Lambda$CDM model
which is currently the most widely accepted cosmological model rests on
the assumption that the universe is statistically isotropic on large scales. This assumption continues to be tested using different observed data such as the cosmic microwave background (CMB)  (see~\cite{Akrami:2019bkn} and references therein) or the distribution of matter (see e.~g.~\cite{Ellis:1984,Singal:2011dy,Secrest:2020has}). 
On the other hand, the statistics that we use to extract information often implicitly assume that the data is isotropic -- for example we typically measure two-point statistics as a function of pairwise separation. In examples where anisotropy is known to be present, such as redshift space distortion, specific corrections can be made and the signal measured. However, to search for unknown anisotropic signals, statistics must be constructed that are agnostic on the underlying structure of the field.

Isotropy is the property of a geometric object to be invariant under rotations. For random fields on a metric space $M$, the property of  isotropy is defined as the covariance function being a function of only the distance between two points on $M$.  One may then ask how this property manifests as a geometrical property of the level or excursion sets of the field. Specifically we ask whether  we can construct a rotationally invariant geometric object using the excursion sets. In two dimensions, given an isotropic field one can intuitively expect that it should be possible to  construct a circle for every excursion set,  such that the radius varies with the threshold field value indexing the excursion set. 
In~\cite{Chingangbam:2017uqv} the contour Minkowski tensor, which is a rank two tensor belonging to the class of morphological descriptors known as Minkowski tensors~\cite{McMullen:1997,Alesker:1999,Hug:2008,Schroder2D:2009}, was used to show the existence of such a series of circles indexed by the field threshold.  
The ratio of the eigenvalues, denoted by $\alpha$,  of this tensor, was then introduced as a statistical tool to test for statistical isotropy using cosmological data.

The basic idea of the test is simple. Exact statistical isotropy implies that $\alpha$ must be unity, which physically means that iso-field boundaries of excursion sets do not exhibit relative alignment. Any alignment will lead to $\alpha<1$, and the closer $\alpha$ is to zero the higher is the {\em degree} of alignment.
In practice, cosmological data is available on spatial regions of finite extent, such as a subset of flat space, or compact unbounded space such as the surface of the sphere. The finiteness of the spatial extent, combined with the resolution of sampling or pixel size, results in relative alignment of the iso-contours that is intrinsic to the sampling. Any real alignment due to a true departure from statistical isotropy will be in addition to this sampling effect. 
In applications, the value of $\alpha$ obtained from a cosmological dataset can be compared with the value obtained from corresponding simulations. In doing so we implicitly assume that the sampling effect in the observed data and the simulations are similar, and can be subtracted. 
This method has been applied to CMB temperature, $E$-mode and lensing convergence data from Planck~\cite{Vidhya:2016,Joby:2018,Goyal:2019vkq,Kochappan:2021fza,Goyal:2021nun}.  

In this paper we revisit how the contour Minkowski tensor (CMT) extracts shape information of arbitrary shaped structures and how it provides a means to sensibly answer questions of the statistical `geometry' of random fields.  In doing so we  extend the results of~\cite{Chingangbam:2017uqv}. 
We first focus on single structures whose boundaries are simple smooth curves and show that the CMT is proportional to the identity matrix if the curve has $m$-fold symmetry, with $m\ge 3$. Then we discuss how the CMT maps any arbitrary curve to an ellipse that is unique up to translations of its centroid. We also carry out a comparison of the shape parameters defined using the CMT with the filamentarity parameter defined using two scalar Minkowski functionals - area and contour length, and discuss the  complementary nature of the shape information that they provide.  
Next, we apply our method to boundaries of excursion sets of random fields. Focusing  on Gaussian isotropic fields we clarify the effect of  finite sampling on the CMT. Using a semi-numerical approach we obtain an analytic expression for the alignment parameter, $\alpha$, which takes into  account the sampling effect. We further obtain an analytic expression for the CMT for Gaussian anisotropic random fields.

Minkowski tensors have used to study a variety of physical effects in astrophysics and cosmology. They have been used to probe the large scale structure of the universe~\cite{Appleby:2018tzk,Appleby:2017uvb,Appleby:2019nit} and the epoch of reionization~\cite{Kapahtia:2017qrg,Kapahtia:2019ksk,Kapahtia:2021eok}. They have also been used to analyse the non-Gaussian nature and statistical isotropy  of Galactic synchrotron emissions~\cite{Rahman:2021azv}, to identity structures in the Large Magellanic Cloud~\cite{Collischon:2021} and to study galaxy shapes~\cite{Rahman:2003,Rahman:2004,Beisbart:2001b}. Apart from astrophysics and cosmology, they have also been applied to a wide variety of physical phenomena such as in condensed matter physics (for e.g.~\cite{Beisbart:2002}) and bio-physics (for e.g.~\cite{Schroder-turk:2018}).

This paper is organized as follows. In section~\ref{sec:sec2} we give a brief overview of smooth random fields and definitions of their symmetry properties. In section~\ref{sec:sec3} we focus on geometric structures in two dimensional space whose boundaries are simple closed curves, define Minkowski tensors and discuss various morphological properties. Section~\ref{sec:sec4} contains our main investigation of the CMT for Gaussian isotropic fields and the effect of finite sampling. In section~\ref{sec:sec5} we discuss the analytic derivation of the ensemble expectation of the contour Minkowski tensor for Gaussian anisotropic fields. We end with concluding remarks in section~\ref{sec:sec6}.  Appendix~\ref{sec:appen_pdfs}
contains the numerical calculation of probability density functions of the anisotropy parameters.  Appendix~\ref{sec:appen_err}
contains the estimation of the numerical error associated with the numerical calculation of the CMT. 

\section{Review of smooth random fields}
\label{sec:sec2}

Let us briefly review the definition of a smooth random field and the symmetry  properties of homogeneity and isotropy.  The definitions follow~\cite{Adler:1981, Yaglom:1986, Adler:2007}. 

Let $M$ be a smooth $n$ dimensional manifold, and let ${\mathbf x}=(x^1,x^2,...,x^n)$ be a coordinate system on a local neighbourhood on $M$. Given a probability space 
 let $f$ be a random variable on the probability space. Consider the family of random variables $f(\vx)$, where each variable is indexed by a point $\vx$ on $M$. Let $f(\vx)$ 
be such that  for any finite $k$ number of points $\vx_1,\vx_2,\ldots,\vx_k$ on $M$, the joint probability density function (PDF), which we denote by ${\cal P}(f(\vx_1),f(\vx_2),...,f(\vx_k))$, of the random variables at those $k$ points is given. Then $f(\vx)$ is called a {\em random field} on $M$.
 
The {\em covariance function} $\xi(\vx,\vx')$ of $f$  is defined as the covariance between the random variables at any two points $\vx$ and $\vx'$,
\begin{equation}
\xi(\vx,\vx') = \bigg\langle \left( f(\vx)- \mu_{\vx} \right) \left( f(\vx')-\mu_{\vx'} \right)  \bigg \rangle, 
  \end{equation}
where  $\mu_{\vx}, \mu_{\vx'}$ are the mean values of the random variables $f$ at $\vx$ and $\vx'$. The auto-covariance gives the variance of $f$ at $\vx$,
\be
\xi(\vx,\vx)=\s^2_{\vx}.
\ee
{\em Homogeneity}:  A random field is said to be (strictly) homogeneous or stationary under a transformation $\vx\rightarrow \vx+{\mathbf  a}$, 
if the joint PDF   
${\cal P}(f(\vx_1),f(\vx_2),...,f(\vx_k))$ is invariant under this transformation. This symmetry  property results in the following consequences that are relevant for our discussion:
\begin{enumerate}
\item The probability distributions of the random variables $f$ at different points on $M$ are identical.
\item $\mu_{\vx}$ and  $\s^2_{\vx}$ are  constant functions on $M$. We will simply denote them as $\mu$ and  $\s^2_0$. 
\item $\xi$ satisfies $\x(\vx,\vx')=\x(\vx-\vx')$. 
\end{enumerate}
So,  the general dependence of $\xi$ on $2n$ coordinate variables is reduced to dependence on $n$ variables only.

\noindent {\em Isotropy}: 
A field $f$ is said to be isotropic if $\xi$ is rotationally invariant, satisfying
\begin{equation}
  \xi(\vx,\vx') = \xi(\Vert \vx-\vx'\Vert),
 \label{eqn:corr_iso} 
\end{equation}
where $\Vert.\Vert$ denotes the geodesic distance on $M$.

\noindent {\em Derivatives of a random field}: If the second partial derivative  $\partial^2\xi/\partial  x^i\partial x'^i$ exists and is finite at the point $\vx=\vx'$, then the first derivative of the field  exists at $\vx$. 
Let $\nabla$ denote the covariant derivative with respect to a connection defined on $M$.   Given that $f$ is a homogeneous field, we can write
\be
\nabla_{x^i} \nabla_{x'^j} \xi(\vx,\vx') =\big\la \nabla_{x^i}f(\vx) \nabla_{x'^j}f(\vx')\big\ra. 
\ee
On the right hand side above the homogeneity of $f$ implies that the derivative of the mean function is zero, and the resulting term is the covariance of the covariant derivatives of $f$. Let us denote $f_{;i}\equiv \nabla_{x^i} f$. The auto-covariance for $\vx=\vx'$   of the components $f_{;i}$ is then denoted by
\be \nabla_{x^i}\nabla_{x^i}\xi(\vx,\vx) = \la f_{;i} f_{;i}\ra  . 
\ee
We can similarly define higher order derivatives. Further, if $f$ is a homogeneous (and isotropic) field, then it follows that its derivatives are also  homogeneous (and isotropic) field.

\noindent {\em Ergodicity}: A  homogeneous field is said to be ergodic if the ensemble expectation can be replaced by spatial average over a realization of the field. 
 \begin{equation}
   \int {\rm d}f\, {\mathcal P}[f] (..) \iff \frac{\int_M \dV  (..)}{\int_M \dV},
 \end{equation}
 where $\dV$ is the infinitesimal volume element. 
 In cosmology it is crucial to assume ergodicity because we have one realisation of the universe.
 
\noindent {\em Gaussian field}: A random field $f(\vx)$ is Gaussian if  ${\cal P}[f(\vx_1),f(\vx_2),...,f(\vx_k)]$ has the form
\be
{\cal P}[f(\vx_1),f(\vx_2),...,f(\vx_k)] = 
\frac{1}{N}
\exp\bigg( -\ha F^T\,\Xi^{-1} F \bigg),
\ee
where $F$ is the array given by
\bea
F&\equiv& \left( f(\vx_1),f(\vx_2),...,f(\vx_k) \right),\\
\Xi_{ij} &=& \xi(\vx_i,\vx_j), \\
 N &=& \sqrt{(2\pi)^k\, {\mathbf{ Det}}\, \Xi},
\eea
with $i,j=1,...,k$.
The derivative of a Gaussian field is also a Gaussian field. Then the field and its derivatives at each point form a set of multivariate Gaussian random fields. We will use this in section~\ref{sec:sec5}.

 \section{Morphology of structures in 2D}
 \label{sec:sec3}

\begin{figure}[H]
\centering\includegraphics[height=1.8cm,width=7.5cm]{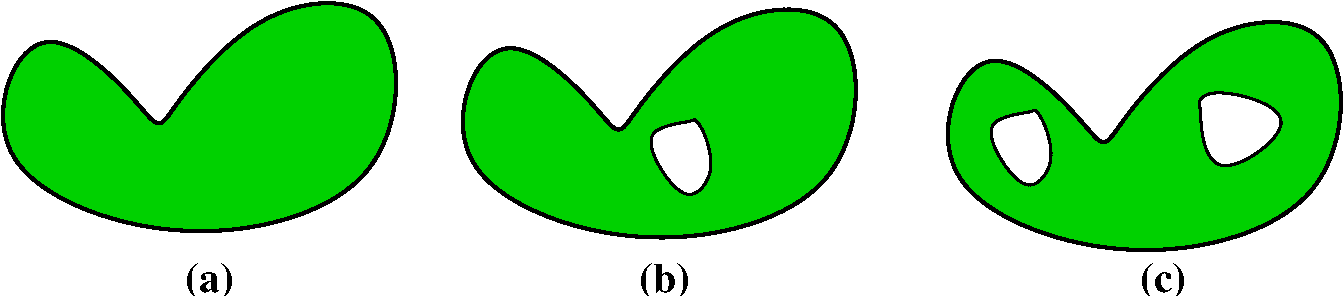}
\caption{Examples of simply, doubly and multiply connected regions. This figure shows that count of structures is equivalent to count of closed curves.}
\label{fig:str}
\end{figure}

 Let us now consider $M$ to be a two-dimensional smooth manifold.  
 Here we are interested in the cases where $M$ is a subset of either flat two-dimensional space $\R2$, or the surface of the sphere $\SS$. 
 We use the word {\em structure} to mean a {\em compact subset} of $M$. The subset can be either a {\em connected region} or a {\em hole}. 
 A {\em simply} connected region has no hole inside it, and has one closed curve as its boundary. A {\em doubly} connected region has one hole inside one connected region. The boundary consists of two closed curves - one that encloses the connected region and another that encloses the hole, and they can be distinguished by assigning different sense of direction. We count a doubly connected region as two structures - one connected region and one hole.  A {\em multiply} connected region consisting of three structures - one connected region and two or more holes.  Fig. \ref{fig:str} shows examples of simply (left panel), doubly (middle panel) and multiply (right panel) connected regions.   
We can generalize the correspondence between a structure and a boundary curve further, and state that {\em each structure can be identified with one unique closed curve.}

 \subsection{Minkowski Tensors for a single structure}
 \label{sec:mt}

Minkowski tensors (henceforth MTs) for a structure, or its associated curve $C$, on  $\mathcal{R}^2$~\cite{Schroder2D:2009} are defined by
\begin{eqnarray}
 W_0^{m}  &=& \int \ {\vec r} \,{}^m \ \ {\rm d}a, \label{eqn:amt}\\
W_1^{m,n} &=& \int_C\ {\vec r} \,{}^m\otimes {\hat n}^n \ \ {\rm d}s, \label{eqn:cmt}\\
W_2^{m,n} &=& \frac{1}{2\pi}\int_C\ {\vec r} \,{}^m\otimes {\hat n}^n \ \kappa\ \ {\rm d}s, \label{eqn:gmt}
\end{eqnarray}
where $\da$ is the area element within the structure, $\ds$ is the infinitesimal arc length on $C$, $\vec r$ is the position vector of each point in the structure, $\hat n$ is the unit normal to the curve, and $\kappa$ is the signed curvature at each point on the curve. $\otimes$ is the symmetric tensor product of two vectors $v$ and $w$ given by $(v\otimes w)_{ij}=(v_i w_j+v_jw_i)/2$. 
Note that the coefficients on the right hand sides of eqn.~\ref{eqn:amt}-\ref{eqn:gmt} differ from earlier papers, e.g. \cite{Schroder2D:2009}. 

The MTs that have rank $m+n=2$ can be classified into translation invariant and covariant ones. There are four translation invariant MTs, of which three are independent in terms of information content. The translation invariance implies that their expressions can be reduced to forms that do not contain $\vec r$. To incorporate this re-expression we denote the three independent translation invariant MTs by $\mathcal{W}_i $, where  $i=0,1,2$.  They are given by, 
\begin{eqnarray}
 \mathcal{W}_0  \equiv W_1^{1,1}  &=&   \int \da\  {\cal I}, \label{eqn:W0} \\
\mathcal{W}_1  \equiv  W_1^{0,2}  &=& \int_C \ {\hat n} \otimes\,{\hat n} \ {\rm d}s,\\
\mathcal{W}_2  \equiv W_2^{0,2} &=& \frac{1}{2\pi} \int_C\ {\hat n} \otimes {\hat n} \ \kappa\ \ {\rm d}s.
\end{eqnarray}
On the right hand side of eqn.~\ref{eqn:W0},  ${\cal I}$ is the identity matrix in two dimensions, and Gauss law has been used to reduce $\W_0$ to the form given on the right hand side~\cite{Hug:2008}. 
$ W_1^{0,2}$ is related to $W_2^{1,1}$, which is the fourth translation invariant MT, by a $90^{\circ}$ rotation. In \cite{Chingangbam:2017uqv} we had used  $W_2^{1,1}$, whereas here we will use $ W_1^{0,2}$ since the expression carries over to higher dimensional manifolds.

The rank zero MTs are the three scalar Minkowski functionals (MFs), $W_i$, where $W_0$ is the area, $W_1$ is the perimeter and $W_2$ is the number of structures. $\mathcal{W}_i $s contain the scalar MFs as their traces:
\bea
    {\rm Trace}\left(\W_i \right) &=&  W_{i}.
\eea  

In \cite{Chingangbam:2017uqv} the $\W_i $ tensors were generalized to structures on curved manifolds and  the special case of the  unit sphere was considered.  In this case invariance under translations on flat space is replaced by invariance under rotations on $\mathcal{S}^2$. 
The  space, $\mathcal{T}_p^*$,  of all possible normal vectors at a point $p\in M$ is isomorphic to the space $T_{p'}^*$ at another point  $p'\in M$. 
The tensor $\hat{n} \otimes \hat{n}$ in the integrands of $\mathcal{W}_1$ and $\mathcal{W}_2$ is an element of the symmetric subspace of the product space $\mathcal{T}_p^*\otimes \mathcal{T}_p^*$, at each $p$.    
Tangent (or cotangent) vectors and elements of product spaces can be summed (or integrated) only when they belong to the same vector space. Implicit in the integral  of $\hat{n} \otimes \hat{n}$ is the geometrical step of parallel transporting all cotangent vectors to one fiducial point on $M$. This step is trivial for flat space and usually not explicitly stated. The translation (or rotational) invariance of $\mathcal{W}_i$ implies that the location of the fiducial point on the curve is not important. In fact, the location of the point on $M$ is not important.
$\mathcal{W}_i$ then transforms as a rank 2 tensor under a {\em local}  rotation $R\in SO(2)$ at the fiducial point $p$. 

We now focus on  $\W_1 $ which we refer to as the {\em contour Minkowski tensor} (henceforth CMT). We can express it as
\begin{equation}
  \mathcal{W}_1  = \left(
  \begin{array}{cc}
    \tau +g_1 & g_2\\
    g_2 & \tau  -g_1
  \end{array}
\right),
\label{eqn:w1_tensor}
\end{equation}
  where
\begin{eqnarray}
  \tau  &=& \frac12 \int_C \,\ds, \label{eqn:tau}\\
  g_1 &=& \frac12 \int_C \, \left( \hat{n}_1^2-\hat{n}_2^2\right)\,\ds, \label{eqn:g1}\\
  g_2 &=& \int_C \, \hat{n}_1\hat{n}_2\,\ds.
\label{eqn:g2}
\end{eqnarray}
$\hat{n}_1,\,\hat{n}_2$ are the components of $\hat{n}$. Since $2\t$ is the perimeter of the curve, $\t$ is always positive, and it is easy to see that $|g_1| < \t$ and  $|g_2| < \t$. 

The right hand side of eqn.~\ref{eqn:w1_tensor} expresses $\W_1$ as a linear combination of the Pauli  matrices, with the coefficient of the complex Pauli matrix $\s_2$ being zero. It   
separates the scalar degree of freedom $\t$ from the pair $(g_1,g_2)$ which captures the true tensorial nature of $\W_1$.  This pair transforms under a local rotation by angle $\theta$ as
\begin{equation}
\left(
  \begin{array}{c}
    g_1\\
    g_2 
  \end{array}
\right) \rightarrow   \left( \begin{array}{c}
  g'_1\\
    g'_2  
  \end{array}\right) =   \left( \begin{array}{cc}
    \cos 2\theta & \sin 2\theta\\
    -\sin 2\theta & \cos 2\theta
\end{array}\right)
\left(   \begin{array}{c}
    g_1\\
    g_2 
  \end{array}\right). 
  \end{equation}
In a given coordinate system, $g_1$ gives a measure of the anisotropic difference between the two components of the normal vectors to the curve, while $g_2$ gives a measure of how correlated the components of the normal vectors are. Using $g_1$ and $g_2$ we can define the scalar quantity $g$ and angle $\varphi$ as
\bea
g &\equiv& \sqrt{g_1^2+g_2^2}, \label{eqn:g}\\ 
\varphi &\equiv& \frac12\tan^{-1}\left( \frac{g_2}{g_1}  \right),\  -\pi/4 \le \varphi < \pi/4.
\label{eqn:phi}
\eea
$g$ represents the magnitude of the tensor while $\varphi$ represents the orientation of elongation or anisotropy with respect to the coordinate system.  

Besides translation invariance, $\mathcal{W}_1 $ is also invariant under parity transformations since all terms are quadratic. Moreover it transforms linearly under size scaling. 

\subsubsection{Relation of $m$-fold symmetry of the curve to rotational symmetry of $\mathcal{W}_1$}
\label{sec:mfold}

The expression for  $\mathcal{W}_1 $ given by eqn.~\ref{eqn:w1_tensor} implies that for it to be proportional to the identity matrix 
we must have $g_1=0$ and $g_2=0$. These conditions must translate into some symmetry properties of the curve. One can then ask the question, what is the class of shapes of the curve that make $\mathcal{W}_1 $ rotationally invariant? It was stated without proof in \cite{Chingangbam:2017uqv} that if the curve has $m$-fold rotational symmetry with $m\ge 3$, then $\mathcal{W}_1$ is invariant under {\em any} local rotation. 

To prove this statement we proceed as follows. 
If the curve has $m$-fold symmetry, then under a local rotation by angle $\delta = 2\pi/m$ (or multiples of $\delta$), we must have
\begin{equation}
  \mathcal{W}_1  \rightarrow  \mathcal{W}_1^{'} = 
  \mathcal{W}_1 .
  \end{equation}
Since $g_1$ and $g_2$ are the components that transform under rotations, if $\mathcal{W}_1 $ remains invariant, we must have $g_1=0=g_2$ under local rotation by $\delta$. Further, if $g_1=0$ and $g_2=0$ in one coordinate system, then they must be zero in any other coordinate system. 
Therefore, if the curve possesses $m$-fold symmetry, then $\mathcal{W}_1 $ is invariant under {\em any arbitrary} rotation. Note that $m$ must be greater than or equal to 3 because $\mathcal{W}_1 $ cannot be invariant for 2-fold symmetry.

\subsubsection{Mapping of a single arbitrary smooth curve to an ellipse, uniqueness and shape anisotropy parameter}
\label{sec:map_ellipse}

In this section we discuss how an arbitrary smooth curve can be mapped to an ellipse. This mapping simplifies the visualization as well as quantification of relative alignments when we study distributions  of arbitrary shaped curves in later sections.   

The eigenvalues of $\mathcal{W}_1 $ for an arbitrary curve are given by
\begin{eqnarray}
  \lambda_1 = \tau -g,\quad
  \lambda_2 = \tau +g.  \label{eqn:l2gt}
\end{eqnarray}
Since $\mathcal{W}_1 $ is real, symmetric and positive definite, the eigenvalues are real and  positive. Inverting  eqn.~\ref{eqn:l2gt} gives
\be
\tau =(\l_1+\l_2)/2, \quad g=(\l_2-\l_1)/2.
\ee

On $\R2$ let us consider the given curve to be an ellipse whose principal axes are $a_1$ and $a_2$, with $a_1<a_2$. Using orthogonal coordinates, let $a_2$ be aligned with the $x$-axis. Then it is straightforward to show that $\mathcal {W}_1$ is diagonal, with the eigenvalues given by
\begin{eqnarray}
\lambda_1 &=& (a_1a_2)^2 \int_0^{2\pi} \frac{\cos^2t}{\left(a_1^2\sin^2 t +a_2^2\cos^2t\right)^{3/2}} \,{\rm d}t, \label{eqn:e1}\\
\lambda_2 &=&  (a_1a_2)^2  \int_0^{2\pi}\frac{\cos^2t}{\left(a_1^2\cos^2t +a_2^2\sin^2t\right)^{3/2}} \,{\rm d}t.
\label{eqn:e2}
\end{eqnarray}
Given  $a_1,a_2$, using these two  equations, we can determine $\l_1,\l_2$.

Conversely, we can ask - given $\l_1,\l_2$ computed for an arbitrary closed curved, can we invert eqns.~\ref{eqn:e1} and \ref{eqn:e2} to determine $a_1,a_2$ uniquely, such that the perimeter remains the same?  The answer is yes. The reason is as follows.  Let  $e=\sqrt{1-a_1^2/a_2^2}$ be the ellipticity of the ellipse, and $E$ be the complete Elliptic integral of the second kind. 
The perimeter of the ellipse is given by $P = 4a_2E(\pi/2,e) = \l_1+\l_2$. 
This provides a relation between $\l_1,\l_2$ and $a_2,e$. Since $E$ is a monotonous function of $e$  there is a one-to-one invertible mapping between $e,P$ and $a_1,a_2$. 
To summarize, the mapping between
\be
(\l_1,\l_2,\varphi) \longleftrightarrow (a_1,a_2,\varphi) \ {\rm or}\ (P, e,\varphi),
\ee
is a one-to-one invertible mapping. Therefore, for any arbitrary simple closed curve there is an ellipse corresponding to it which is unique up to the location of its centroid. 

{\em $\W_1$ for open curves}:  $\W_1$ is well defined as a tensor even when the curve is not closed. It is still independent  of which fiducial point on $M$ is used for computing it, and so we can use the above mapping to construct an ellipse corresponding to any smooth open curve. In fact, we can go further and state that given any real, symmetric and positive definite $2\times 2$ matrix, using this mapping we can construct an ellipse which is unique up to translations of its centroid.  

{\em Generalization of the correspondence between arbitrary curve and ellipse to curves on the unit sphere}: 
On the unit sphere, an ellipse may be defined as the locus of points for which the sum of the geodesic distances from the two loci is constant. It is straightforward to generalize the above arguments from flat 2D space to the unit sphere.

\begin{figure}
\includegraphics[height=6cm,width=6cm]{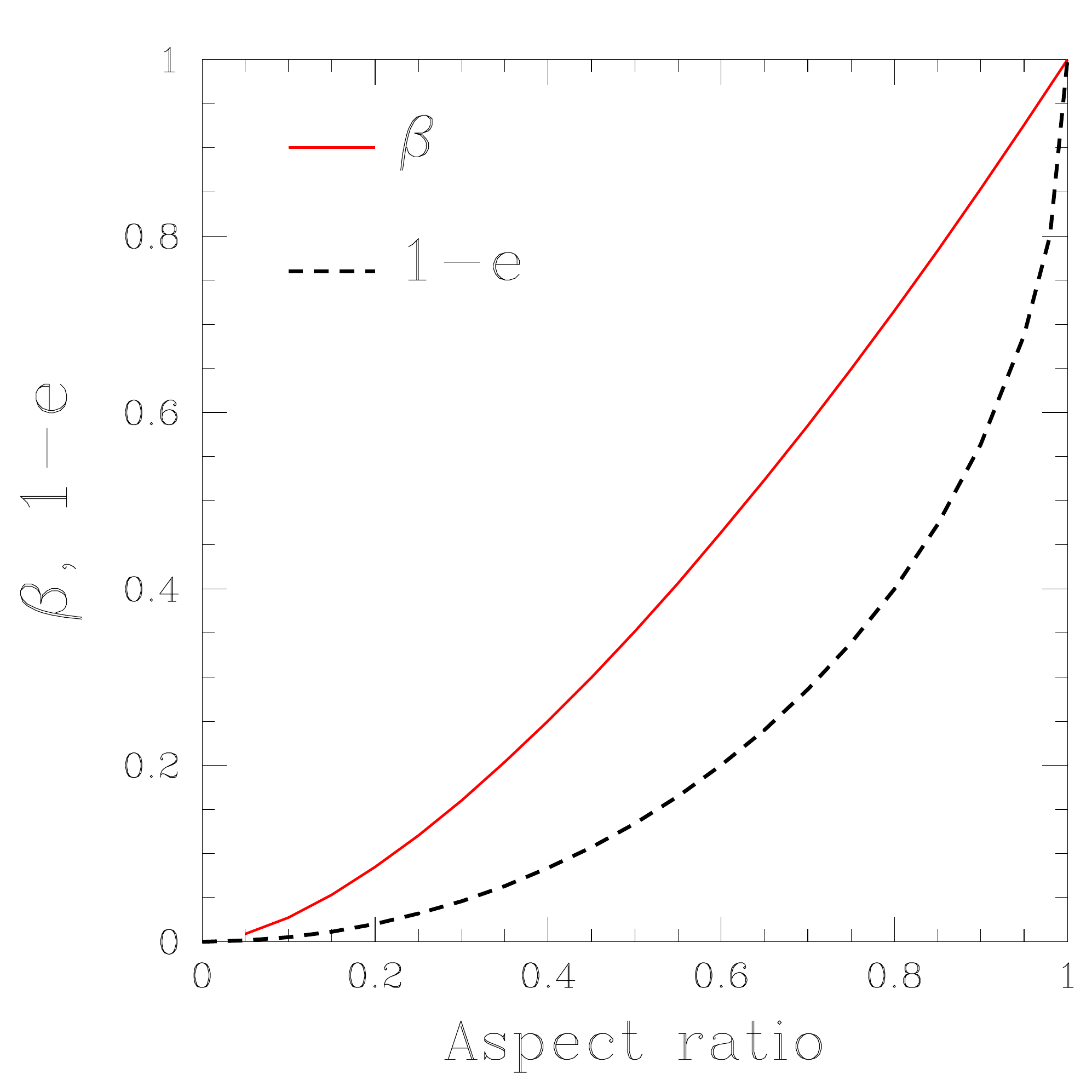}
\caption{$\beta$ versus the aspect ratio, $a_1/a_2$ of an ellipse. The function $1-e$, $e$ being the ellipticity, is also shown for comparison.} 
\label{fig:b}
\end{figure}

{\em Anisotropy parameter}: Next, to quantify the anisotropy of a curve, let  us define the quantity $\beta$ to be the ratio of the eigenvalues~\cite{Schroder2D:2009}, 
\begin{equation}
\beta  \equiv\frac{\lambda_1}{\lambda_2} =  \frac{\tau  - g}{\tau +g}.
\end{equation}
 The value of $\beta$ lies between zero and one.  From the result of the previous subsection we get $\beta$  equal to one for a closed curve having $m$-fold symmetry, with $m\ge 3$. Deviation of  $\beta$ from one, or $g$ from zero indicates anisotropy of the curve. Fig.~\ref{fig:b} shows how $\beta$ varies with increase of the aspect ratio $a_1/a_2$ of ellipses on $\R2$. The function $1-e$, where $e$ is the ellipticity, is also shown for comparison. 


\subsection{Many curves and their relative alignment}
\label{sec:cmt_many}

We now discuss $\W_1$ for a collection of many arbitrary shaped smooth simple curves. 
For simplicity let us first consider two closed curves $C'$ and $C''$ whose CMTs are  $\mathcal{W}^{'}_1$ and $\mathcal{W}^{''}_1$, respectively. Let their tensor  sum be 
$\widetilde{\mathcal{W}}_1  \equiv \mathcal{W}^{'}_1 + \mathcal{W}^{''}_1$ is   
\begin{eqnarray}
  \widetilde{\mathcal{W}}_1  &= & \left(
  \begin{array}{cc}
    \tau +g_1 & g_2 \\
    g_2 & \tau -g_1
  \end{array}
  \right),
  \end{eqnarray}
where
\begin{eqnarray}
  \tau =   \left(  \tau'+\tau'' \right),\quad
  g_1 &=&  \left(g'_1+g''_1\right),\quad
  g_2 =  \left(g'_2 + g''_2\right), \nn\\
  \varphi &=& \frac12\tan^{-1}\left(\frac{g_2}{g_1}\right).
  \end{eqnarray}

\begin{figure}
  \centering \includegraphics[height=7.cm,width=8.cm]{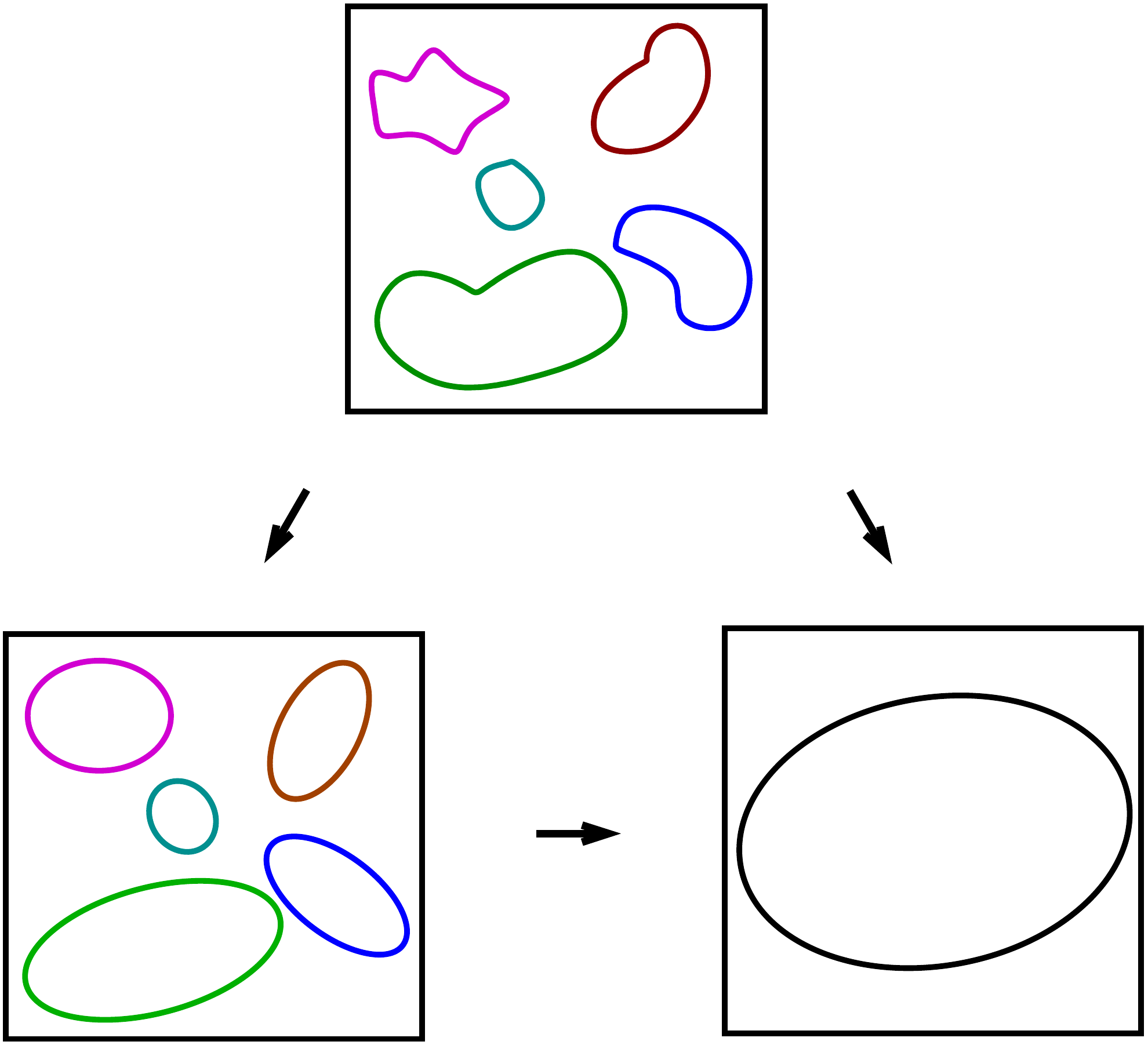}\\
  \centering a \hskip 4cm b 
  \caption{Schematic diagram for mapping from a distribution of curves (top panel) to a distribution of ellipses (a),
    and then to a single ellipse (b). The direction of the arrows indicate that we can obtain the final ellipse in panel (b) with or without first mapping each individual curve to an ellipse. }  
\label{fig:manyellipses}
  \end{figure}

Using the mapping between eigenvalues and principle axes of an ellipse described earlier we can  construct the ellipse that corresponds to $\widetilde{\mathcal{W}}_1 $. 
It is then straightforward to generalize the mapping to a distribution of many curves. Fig.~\ref{fig:manyellipses} shows a schematic diagram showing mapping of many curves to a final single ellipse.

{\em Alignment parameter}: Given a spatial distribution of many curves the relative alignment between the curves is encoded in  the parameter~\cite{Schroder2D:2009},
\begin{equation}
  \alpha \equiv\frac{\Lambda_1}{\Lambda_2},
\end{equation}
where $\Lambda_1$ and $\Lambda_2$ are the eigenvalues of  ${\widetilde{\mathcal{W}}}_1 $ such  that $\Lambda_1\leq\Lambda_2$. By definition we have $0\leq\alpha \leq1$. $\alpha $ gives a measure of the deviation from rotational symmetry in the spatial distribution of structures. We obtain $\alpha  = 1$ if there is no preferred orientation in the arrangement of the structures. For example, two identical ellipses placed such that their semi-major axes form $90^{\circ}$ between them gives $\alpha  = 1$. For $\alpha<1$, its value gives the {\em degree of anisotropy}, while the orientation information is given by $\varphi$ obtained from ${\widetilde{\mathcal{W}}}_1 $. 
For a single curve we have $\alpha =\beta$. 

We can expand $\alpha $ in terms of $g/\t$ as
\begin{equation}
  \alpha  = \frac{\tau - g}{\tau+g} = 1 - 2\frac{g}{\tau} + 2\frac{g^2}{\tau^2} -
  \mathcal{O}\left( \frac{g^3}{\tau^3}\right).
  \label{eqn:alpha-gtau} 
\end{equation}
This expression for $\alpha$ shows that there can be degeneracy between the number of structures and the total perimeter, in the way $\alpha$  captures the information of alignment. Any disproportionate change of $\tau$ and $g$ will make $\alpha$ change, with the shift either towards one or towards zero determined by increase or decrease of their ratio.

Note that $g$, or $g/\t$, or $(\a,\beta)$ are equivalent measures of intrinsic anisotropy.  Which of these three is best suited for statistical analysis will be determined by the size of the standard deviation when we apply to random fields. We will see in section~\ref{sec:sec5} that the statistical fluctuations of $\alpha$  are considerably smaller than that of $g$. This is due to cancellation of the fluctuations when taking the ratio of the eigenvalues (see fig.~3). Hence $(\a,\beta)$ are better suited as  anisotropy parameters. 
\subsection{Comparison with shape finders}
\label{sec:cmt_shape}

Using the scalar MFs, $W_0$ and $W_1$ -- normalised such that they correspond to area and perimeter respectively -- one can define the {\em filamentarity} parameter~\cite{Sahni:1998cr,Bharadwaj:1999jm}:
\begin{equation}
\label{eq:fil} F(\nu) = \frac{W_1^2-4\pi W_0 }{W_1^2+4\pi W_0 }. 
\end{equation}
$F$ provides a measure of the morphology of a structure, specifically its filamentarity. By definition its value lies between 0 and 1, and takes extremal values $F = 0$ for a spherical disc and $F=1$ for a filament of vanishing thickness. It is of interest to compare how the anisotropy of structures manifests in $F$ and $\alpha$ (or $\beta$ for individual structures). To do so we have taken two simple examples of sums of Gaussian functions given by
\be
f(x,y)= \sum_{i=1}^n\exp\left\{-\frac{ (x-x_i)^2 + (y-y_i)^2}{2\s_i^2}\right\},
\label{eqn:gf}
\ee
such that the Gaussian peak locations  $x_i,y_i$ and widths $\s_i$ are chosen depending on how we want to arrange the iso-contours.

\begin{figure}
 \hskip .6cm \includegraphics[height=3.6cm,width=3.6cm]{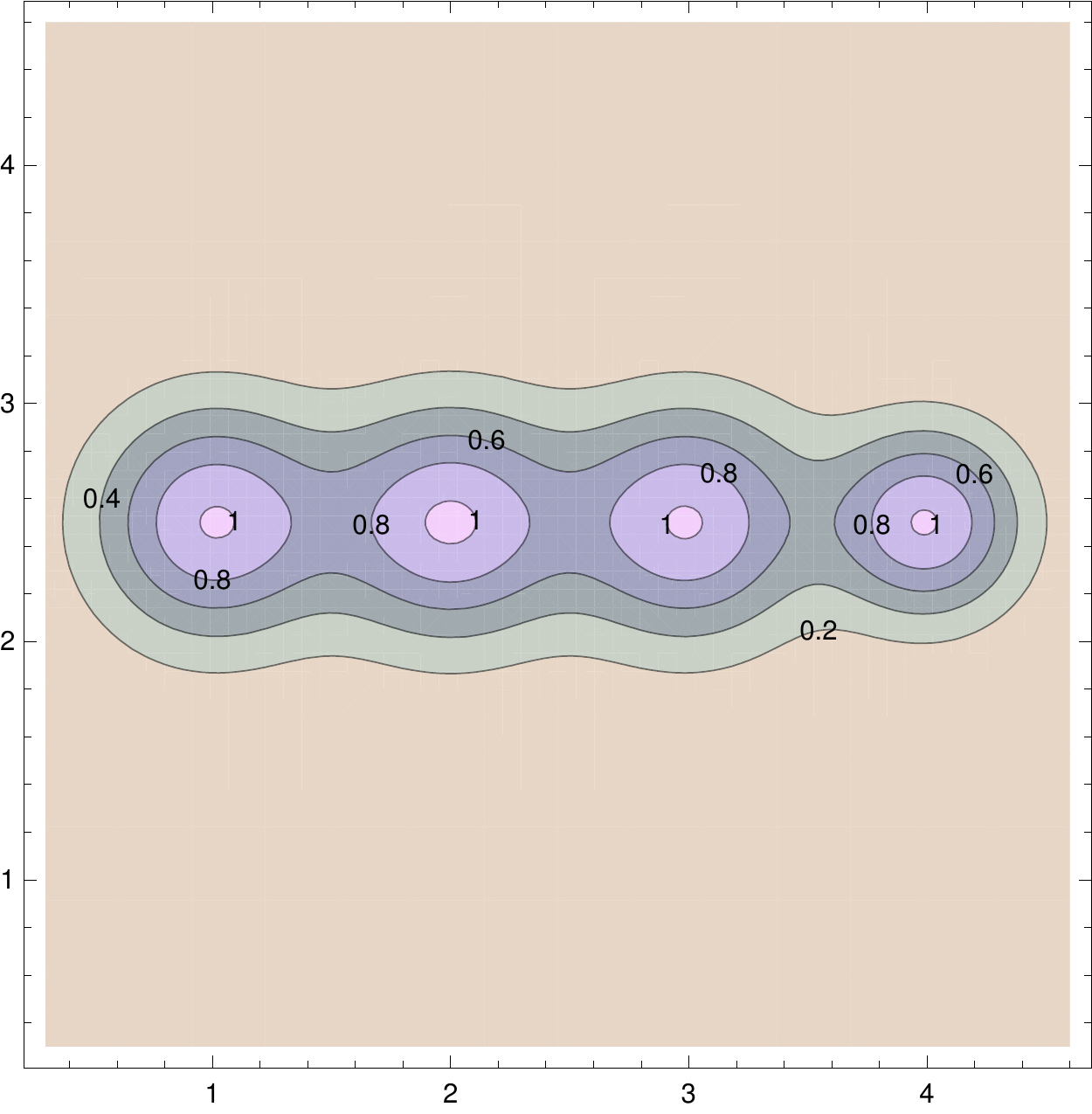}\hskip .7cm
  \includegraphics[height=3.6cm,width=3.6cm]{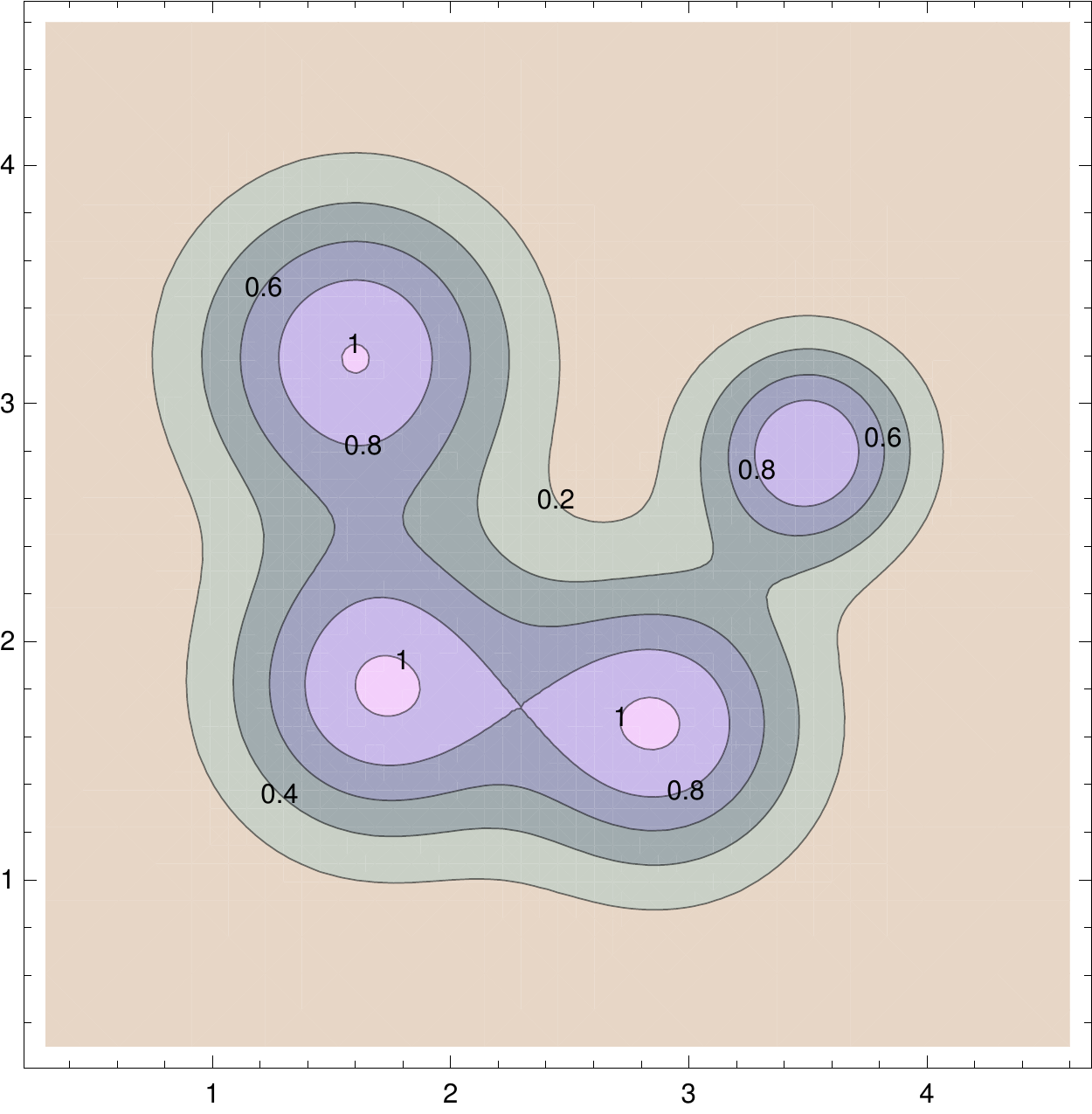}\\ 
  \includegraphics[height=4.2cm,width=4.2cm]{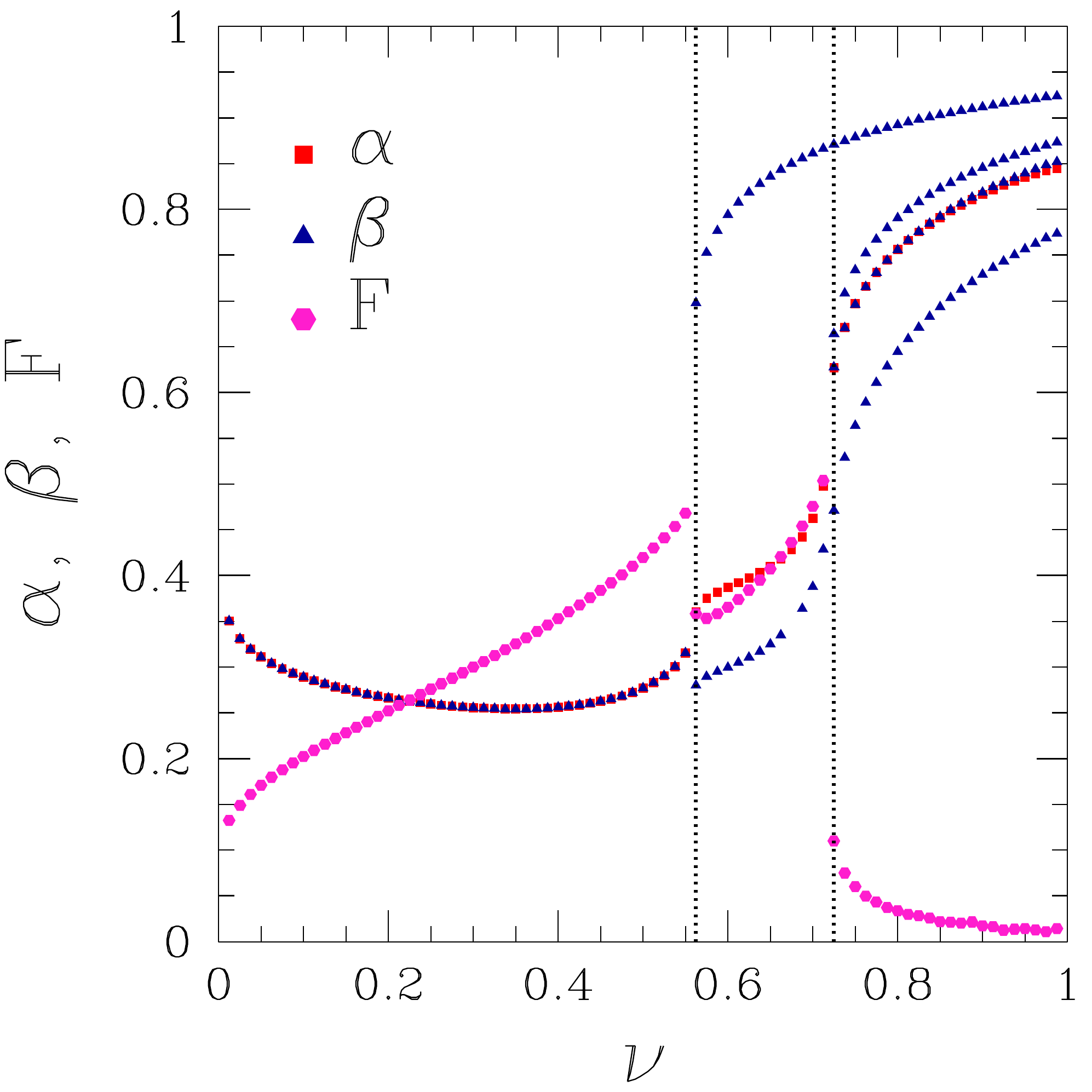}
  \includegraphics[height=4.2cm,width=4.2cm]{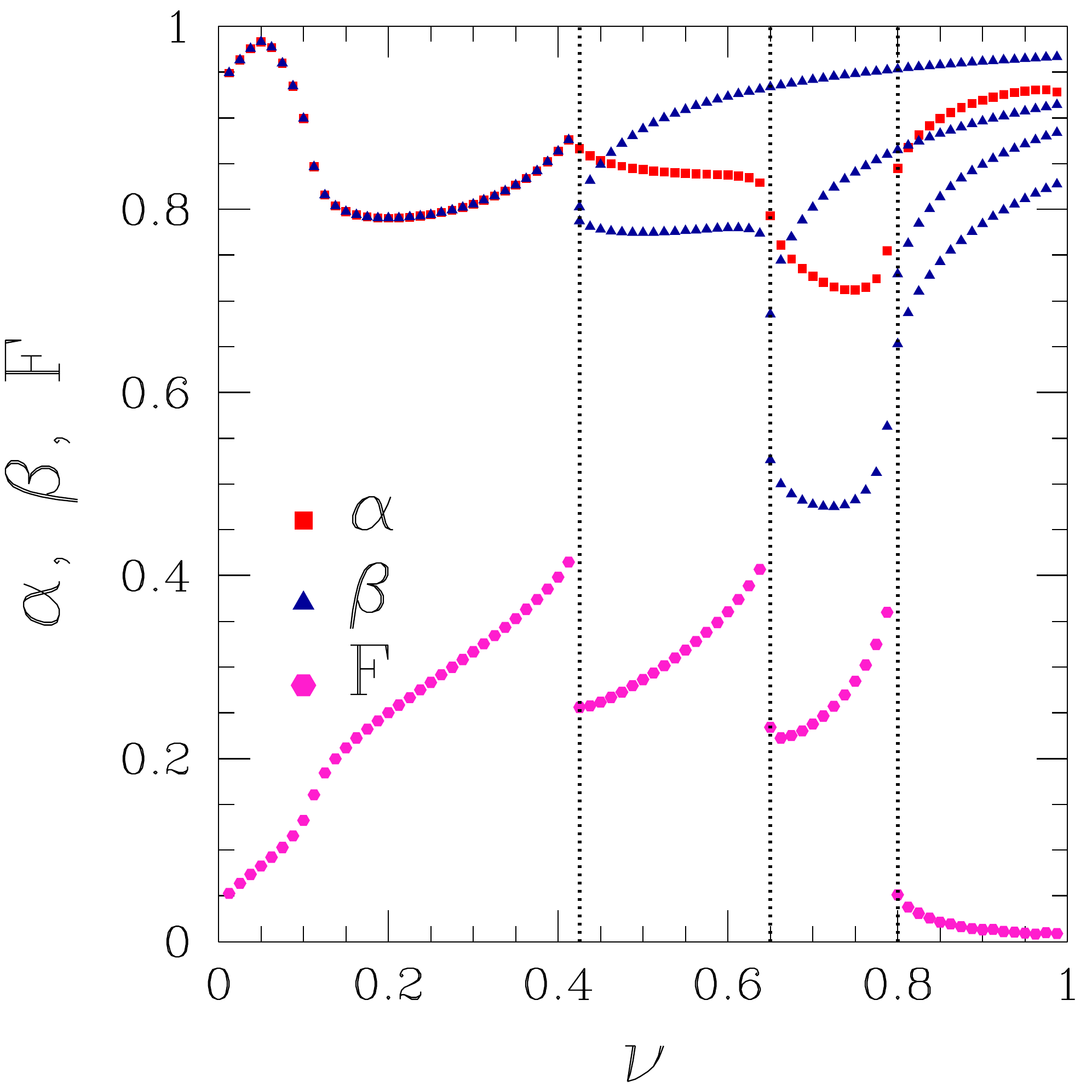}
  \caption{{\em Top}: Iso-contours for two functions given by eq.~\ref{eqn:gf} with peak locations arranged linearly (left) and non-linearly (right). 
    {\em Bottom}: $F$, $\alpha$ and $\beta$ for individual structures after fragmentation versus threshold, for the corresponding function in the panel above.  The percolation thresholds are indicated by the black dashed lines.}
\label{fig:compare_a_F}
\end{figure}
The top panels of Fig.~\ref{fig:compare_a_F} show two examples of $f$ where $x_i,y_i$ are arranged linearly (left) and curvilinearly (right). 
The bottom panels show $\alpha$, $\beta$ and $F$ for the function in the corresponding panel above. The thresholds at which the iso-contours fragment are marked by the dotted lines. When the excursion sets fragment, we define $F$ as the linear sum of ($\ref{eq:fil}$) obtained from each individual structure.

We find for both cases that $F$ is low for $\nu \sim 0$ but increases as the contours shrink, as the area decreases relative to the perimeter. $F$ detects elongation of structures but is not sensitive to whether the overall structure is linear or curved. At each fragmentation point, the function exhibits discontinuity. At high thresholds $\nu \sim 1$, the contours reduce to four circular peaks and $F$ approaches zero. $F$ is qualitatively similar for the two fields presented in the Figure. The statistic $\alpha$, on the other hand, is  strongly sensitive to the manner in which the iso-contours change. Therefore, it contains more information of the morphology of the curve compared to $F$. However, the caveat is that it will not sensitively distinguish a structure which is, say, mildly elliptical from one which is elongated but curved, such as the example on the right panel of fig.~\ref{fig:compare_a_F}. Therefore, for practical applications it will be best to use a combination of both $F$ and $\alpha$. One important distinction between the two statistics is that the CMT provides directional information, encoded in $\varphi$ (eqn.~\ref{eqn:phi}) which cannot be obtained from the scalar MFs. 

It is also interesting to observe the behaviour of $\beta$ for individual structures. As we can see in the bottom panels of fig.~\ref{fig:compare_a_F}, till the first fragmentation there is only structure and so $\alpha=\beta$. After each fragmentation threshold the blue curve for $\beta$ bifurcates into two curves for each individual structure. Each blue curve approaches unity as the excursion sets reduce to a set of circular perimeters enclosing the peaks at $\nu \sim 1$.

\section{Statistical isotropy of smooth random fields }
\label{sec:sec4}

Having discussed in the previous section the morphology of individual structures and spatial distribution of many structures, we are now equipped to discuss smooth random fields. Our focus here is on clarifying the meaning of statistical isotropy geometrically and to quantify the effect of finite resolution  on $\alpha$.

Let $f$ be a Gaussian field whose mean and standard deviation are $\mu$  and $\s_0$, respectively. Let $\s_1$ be defined as 
\be
\s_1^2\equiv \langle f_{;1}^2 \rangle + \langle f_{;2}^2 \rangle = \s_{f_{;1}}^2 + \s_{f_{;2}}^2. 
\ee
We will work with the mean subtracted and normalized field $u\equiv (f-\mu)/\s_0$. 
For a chosen threshold value of $u$, denoted by $\nu$, the set of all points on $M$ where $u>\nu$ is referred to as an {\em excursion} or {\em level} set, denoted by $\mathcal{Q}_{\nu}$. It consists of a set of connected regions, each of which may be simply or multiply connected with holes in them. The boundary of $\mathcal{Q}_{\nu}$, denoted by $\partial\mathcal{Q}_{\nu}$, consists of closed iso-threshold contours that enclose connected regions and holes.
\subsection{Contour Minkowski tensor for random fields}

Let $\widetilde{\mathcal {W}}_1$ be  the sum over the individual  $\W_1$ of each  curve at a threshold $\nu$,
In order to express $\widetilde{\mathcal {W}}_1$ in terms of the field, we use 
$\hat{n}_i = \frac{u_{;i}}{\left| \nabla u \right|}$. 
Then,  we get
 \begin{eqnarray}
  \left( \widetilde{\mathcal {W}}_1\right)_{ij} &=& \frac{1}{V}\int_{{\partial \mathcal Q}_{\nu}} \,{\rm d}s \, \frac{u_{;i} \,u_{;j}}{\left| \nabla u \right|^2}.
  \label{eqn:w1u}
 \end{eqnarray}
Note that we have introduced a factor $1/V$ in the definition above in comparison to the expression in section~\ref{sec:mt}. By taking $V$ to be proportional to the volume of $M$ (which is the area in this case), 
$V= 4\times {\rm Area\ of}\ M$, 
we can relate the trace of $\widetilde{\W}_1$ to the standard definition of the contour length MF in cosmology. Here we will use $V= {\rm Area\ of}\ M$.
 
For explicit calculations, both numerical and analytic,  it is convenient to express the integral over $\partial\mathcal{Q}_{\nu}$ into area integral over the entire manifold $M$ by introducing a Jacobian and  delta function $\delta(u-\nu)$ constraint, as
\begin{eqnarray}
   \left( \widetilde{\mathcal {W}}_1\right)_{ij} &=&  \frac{1}{V}\int_{M} \da \, \, \delta(u-\nu)\ \frac{u_{;i} \,u_{;j}}{|\nabla u|},
  \label{eqn:w1_area}
 \end{eqnarray}
where $\da$ is the area element. 
The expressions for $g_1$, $g_2$ and $\tau$ in terms of the field derivatives are then given by
 \begin{eqnarray}
  \tau &=& \frac{1}{2V} \int_M \da  \ \delta(u-\nu)\ |\nabla u|, \label{eqn:tau_u}\\
  g_1 &=& \frac{1}{2V}\int_M \da  \ \delta(u-\nu)\ \frac{u_{;1}^2-u_{;2}^2}{|\nabla u|},  \label{eqn:g1_u}\\
  g_2 &=& \frac{1}{V}\int_M  \da  \ \delta(u-\nu)\ \frac{u_{;1}u_{;2}}{|\nabla u|}.
 \end{eqnarray}

\noindent and we also define $g = \sqrt{g_{1}^{2} + g_{2}^{2}}$. To numerically compute $\widetilde\W_1$ we follow a method that was first put forth in~\cite{Schmalzing:1997aj} and adapted in~\cite{Schmalzing:1998,Chingangbam:2017uqv}. 
It involves expressing the area integral of $\widetilde\W_1$ as a sum over equal area pixels on $M$.  Let  $\Delta \nu$ denote the threshold bin size. Then the $\delta-$function can be approximated as 
 $\delta \left( u-\nu \right) = 
    1/\Delta \nu, \quad {\rm if} \ 
   u \in \left( \nu-\frac{\Delta \nu}{2} , \nu+\frac{\Delta \nu}{2} \right)$, and zero otherwise. 
  Let the total number of pixels be $N_{\rm pix}$.   The estimator for  $\widetilde\W_1$ can then be expressed as 

\be
   \left(\widetilde\W_1\right)_{ij} = \frac{1}{N_{\rm pix}}\sum_{k=0}^{N_{\rm pix}} \, \frac{w_k}{\Delta\nu}  \ \frac{u_{;i}(k) u_{;j}(k)}{|\nabla u(k)|}.
 \ee
 The variable $w_k$ is the weight factor of the delta function for the $k$th pixel. It has value 
 one if the field at $k$ pixel has value in the range $\nu-\Delta\nu/2$ to $\nu+\Delta\nu/2$, and zero otherwise. In case there is masking of parts of the manifold $M$, $N_{\rm pix}$ denotes the total number of unmasked pixels.

Let $\langle .\rangle$ denote ensemble averaging. 
The ensemble expectation of $\widetilde{\mathcal{W}}_1$ gives,
\begin{equation}
\langle  \widetilde{\mathcal{W}}_1 \rangle  =  \left(  \begin{array}{cc}
  \langle\tau\rangle  +\langle g_1\rangle & \langle g_2\rangle\\
    \langle g_2\rangle &   \langle\tau\rangle  -\langle g_1\rangle
  \end{array}
  \right).
\label{eqn:w1_ens}
\end{equation} 
Then $\alpha$, denoted with an overbar,  computed from $\langle  \widetilde{\mathcal{W}}_1 \rangle$ is 
\be
\overline\a = \frac{\la \tau\ra - \sqrt{\langle g_1 \rangle^2 + \langle g_2 \rangle^2 }}{\la \tau\ra + \sqrt{\langle g_1 \rangle^2 + \langle g_2 \rangle^2 }}.
\label{eqn:a1}
\ee
Alternatively, we can first diagonalize  $\widetilde{\mathcal{W}}_1$ for each  realization, with the convention that the (11) element is less than the (22) element, and then carry out the ensemble averaging. Geometrically the diagonalization means we rotate the ellipse obtained for each realization of the field such that its axes are aligned with the coordinate system. And choosing the (11) element to be  less than the (22) element means the rotation is by angle $\pi-\varphi$, using the right hand convention.  $\varphi$ is given by eq.~\ref{eqn:alpha-gtau}. We denote the rotated $\widetilde{\mathcal{W}}_1$ by $\widetilde{\mathcal{W}}_1^R$. 
 $\varphi$ is a random angle and will be different for different realizations in general. 
Then the ensemble expectation of the corresponding $\a$ is
\be
\la\a\ra= \left\la \frac{\t-g}{\t+g}\right\ra .
\label{eqn:a2}
\ee
In general, eqns.~\ref{eqn:a1} and \ref{eqn:a2} give different answers. 

For comparing between observation and simulations eqn.~\ref{eqn:a1} is not appropriate since $g_1,\,g_2$ depend on the coordinate choice. The coordinate independent  way to compare is to use eqn.~\ref{eqn:a2}. If $g$ and $\t$ are uncorrelated, as they should be since they are independent degrees of freedom, 
we get 
\be
\la\a\ra\simeq 1 - 2\frac{\langle g\rangle}{\langle \tau \rangle} + \mathcal{O}\left( \frac{\la g\ra ^2}{\la\tau\ra^2}\right). 
\label{eqn:a2series}
\ee

In \cite{Chingangbam:2017uqv} it was shown for Gaussian and Rayleigh isotropic fields that the ensemble expectation of $\widetilde{\mathcal {W}}_1$ is proportional to the identity matrix. The reason is that statistical isotropy implies $\la g_1\ra = 0 = \la g_2\ra$. As a consequence,  $\alpha$ given by eqn.~\ref{eqn:a1} is one at every threshold. In fact, {\em this will be true in general for any isotropic field, regardless of its PDF.} 
This result is however not useful for comparison between observed data and simulations, as mentioned above. What we need instead is an analytic expression for eqn.~\ref{eqn:a2} or \ref{eqn:a2series}. 

In practical applications where data is available over  finite spatial extent, and finite pixel size, $g_1, g_2,g$ will be non-zero even though the field is given to be isotropic. This is due to statistical fluctuations caused by the sampling. 
We expect $\la g_1\ra$, $\la g_2\ra$ and $\la g\ra$ to carry  characteristic information of the finite sampling and this is what we investigate in the next subsection.


\subsection{Quantifying the effect of finite sampling and resolution}  
\label{sec:sec4B}

In any realistic calculation there are two inherent length scales - the size of $M$, which can be either a box on flat space or the sky region on the sphere, denoted by $L$, and the typical size of structures, which we denote by $l_s$. Since the pixel size is typically smaller than $l_s$ we need not consider it as another scale. $l_s$ is determined by a convolution of the inherent correlation length of the physical interaction(s) of the stochastic process that generate the field, and any smoothing applied to the field. Here we consider smoothing using a Gaussian kernel. Then we can define the {\em finite sampling} parameter $s$  to be:
 \be
s\equiv\frac{{ l_s}}{L}.
 \ee
 $s$ quantifies the effect of finite sampling of the field which is determined by $L$ and $l_s$. Note that $s$ can approach zero in two ways - $l_s$ approaching zero, or $L$ approaching infinity. 
 
 We will work with Gaussian isotropic CMB temperature simulations generated using \texttt{Healpix}~\cite{Healpix}, with the input angular power spectrum obtained from \texttt{CAMB}~\cite{cambsite}. 
 Here {\em isotropy} refers  to the fact that the input angular power spectrum $C_{\ell}$ for the simulated maps  
 is obtained from a primordial power spectrum, $P(k)$, which depends only on the amplitude $k$ of the wave modes. 
 The maps have input cosmological parameters given by Planck~\cite{Planck:2015}.  
 $P(k)$ has the functional form $P(k)=A_s\left(k/k_0\right)^{n_s-1}$, where $A_s$ and $n_s$ are the amplitude and spectral index, respectively, of the primordial perturbations, and $k_0$ is the pivot scale. 
 
 Since $n_s$ is close to one the field has  fluctuations (hence the excursion sets have structures), at all scales. Smoothing has the effect of introducing a cut-off of the fluctuations at the scale set by the smoothing scale. The variation of $l_s$ is therefore determined by the variation of the smoothing scale.       
We use the full sky for our calculations here, and so $L$ is a constant given by the surface area of a unit sphere, $L^2=4\pi$. Consequently, the variation of $s$ is determined by the variation of the smoothing scale. We denote the smoothing scale by $\theta_s$ which we take to be the same as the full width at half maximum (FWHM).  
 
We focus on the following two questions. First, for a given $s$, what are the functional forms of the ensemble mean values $\la g\ra $ and $\la\t\ra$ as functions of threshold values?  Secondly, how do the functional forms of $\la g\ra$ and $\la \t\ra$ scale or vary with $s$?  
For $\la\t\ra$, the answers to both these questions for a Gaussian field having {\em exact} isotropy symmetry and input power law power spectrum, are well known. Its functional dependence on $\nu$ is given by~\cite{Tomita:1986},
\be
\la \tau \ra^{\rm exact}(\nu)\propto \frac{\s_1}{\s_0} e^{-\nu^2/2}. 
\label{eqn:tau_gauss}
\ee
The dependence on the smoothing scale is encoded in the correlation length $r_c={\s_1}/{\s_0}$, which scales roughly as $r_c \propto \theta_s^{-1}$~ \cite{BBKS:1985,Bond:1987ub}. 
We can expect that the anisotropy of the field introduced by finite sampling will modify eqn.~\ref{eqn:tau_gauss}.  
For $\la g(\nu)\ra$, however, we don't know the answers to the above questions and we will answer them  using numerical computations. 

\begin{figure}
\centering\includegraphics[height=10cm,width=8.7cm]{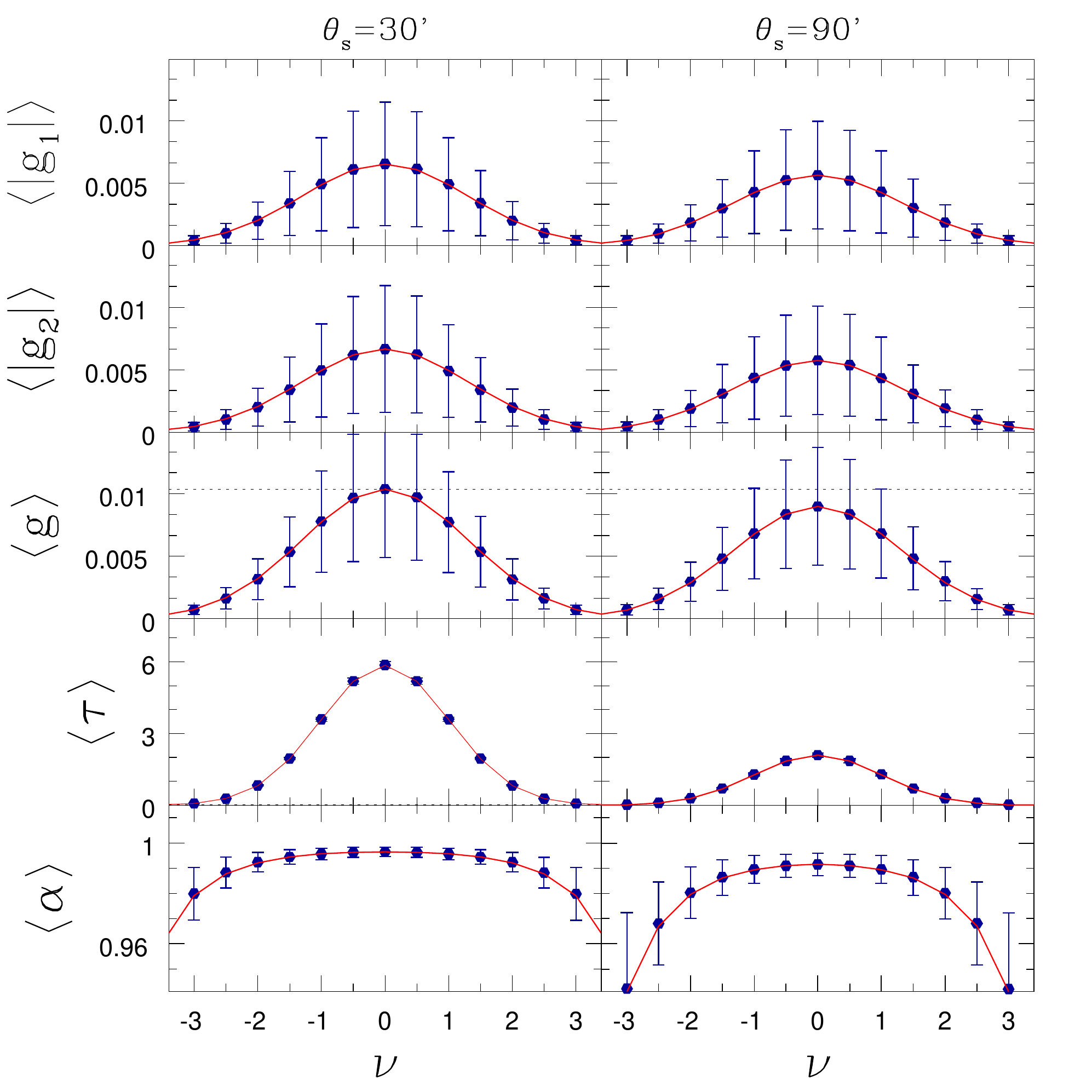}\\
\caption{The blue dots are mean values of $|g_1|$ (top row), $|g_2|$ (second row), $g$ (third row), $\tau$ (fourth row), and $\alpha$ (bottom row),  calculated from $10^4$ Gaussian isotropic simulations for two smoothing scales. For $|g_1|$, $|g_2|$, $g$ and $\tau$, the red solid lines are Gaussian fitting functions. 
For $\alpha$, the red solid line is the corresponding function defined by eqn.~\ref{eqn:alphamean}. The error bars are the standard deviations obtained from the $10^4$ simulations.}   
\label{fig:2d_allnu}
\end{figure}

We numerically calculate $g_1,g_2,g,\t,\alpha$ for different values of $\theta_s$ from $10^4$ simulated CMB temperature maps.   
We use threshold bin size $\Delta\nu=0.5$.  The blue dots in Fig.~\ref{fig:2d_allnu} show the numerically computed values of  $\la|g_1|\ra$ (top row), $\la|g_2|\ra$ (second row), $\la g\ra$ (third row), $\la\tau\ra$ (fourth row) and $\la\a\ra$ (bottom row) versus $\nu$ for two different smoothing scales.  The error bars are the standard deviations obtained from the simulations. 

{\em Functional forms of $\la|g_1|\ra$, $\la|g_2|\ra$ and $\la g\ra$}: It can be discerned from the panels in the top two rows of  fig.~\ref{fig:2d_allnu} that $\la |g_1|\ra$ and $\la |g_2|\ra$ can be fit well by  Gaussian functions of $\nu$, which we write as,
 \bea
\langle |g_1(\nu)|\rangle &=& A_{g_1} e^{-\nu^2/2\sigma_{g_1}^2}, \label{eqn:g1mean}\\
\langle |g_2(\nu)|\rangle &=& A_{g_2} e^{-\nu^2/2\sigma_{g_1}^2}. \label{eqn:g2mean}
 \eea
It is also clear that $\la |g_1|\ra \simeq \la |g_2|\ra$, and so $A_{g_1} =A_{g_2}$ and $\s^2_{g_1} =\s^2_{g_2}$. The red solid lines in the figure correspond to fits to these Gaussian functions. 

 Next, using $A_{g_1} =A_{g_2}$ and $\s^2_{g_1} =\s^2_{g_2}$ we get $\la g\ra$ as 
\bea
\langle g(\nu)\rangle &=& A_g e^{-\nu^2/2\sigma_g^2}, \label{eqn:gmean}
\eea
where $A_g=\sqrt{2} A_{g_1}$ and $\s_g=\s_{g_1}$. The red solid lines in the panels showing $\la g\ra$ correspond to the fits to eqn.~\ref{eqn:gmean}. 

We can also define an angle $\widetilde\varphi$  
\be  \frac{\la |g_2|\ra}{\la |g_1|\ra} = |\tan 2\widetilde\varphi| ,\ee
 where we have put a tilde over $\widetilde\varphi$ on the right hand side to distinguish it from $\varphi$ which is defined without the modulus on $g_1$ and $g_2$. Using $\la |g_1|\ra = \la |g_2|\ra$ we get 
 \be
 \widetilde\varphi = \pi/8.
 \ee
 
{\em Functional form of $\la\tau\ra$}:   For $\la\t\ra $ the Gaussian fitting function is 
 \be
\la \tau(\nu) \ra = A_{\t} e^{-\nu^2/2\s_{\t}^2}.  \label{eqn:taumean}
 \ee
 The difference between this equation  and eqn.~\ref{eqn:tau_gauss} is that we have introduced $\s_{\t}$ which can be different from one, due to the sampling effect. The red solid lines in the panels showing $\la \t\ra$ again correspond to the fit to eqn.~\ref{eqn:taumean}. 

\begin{figure}
\includegraphics[height=6cm,width=6cm]{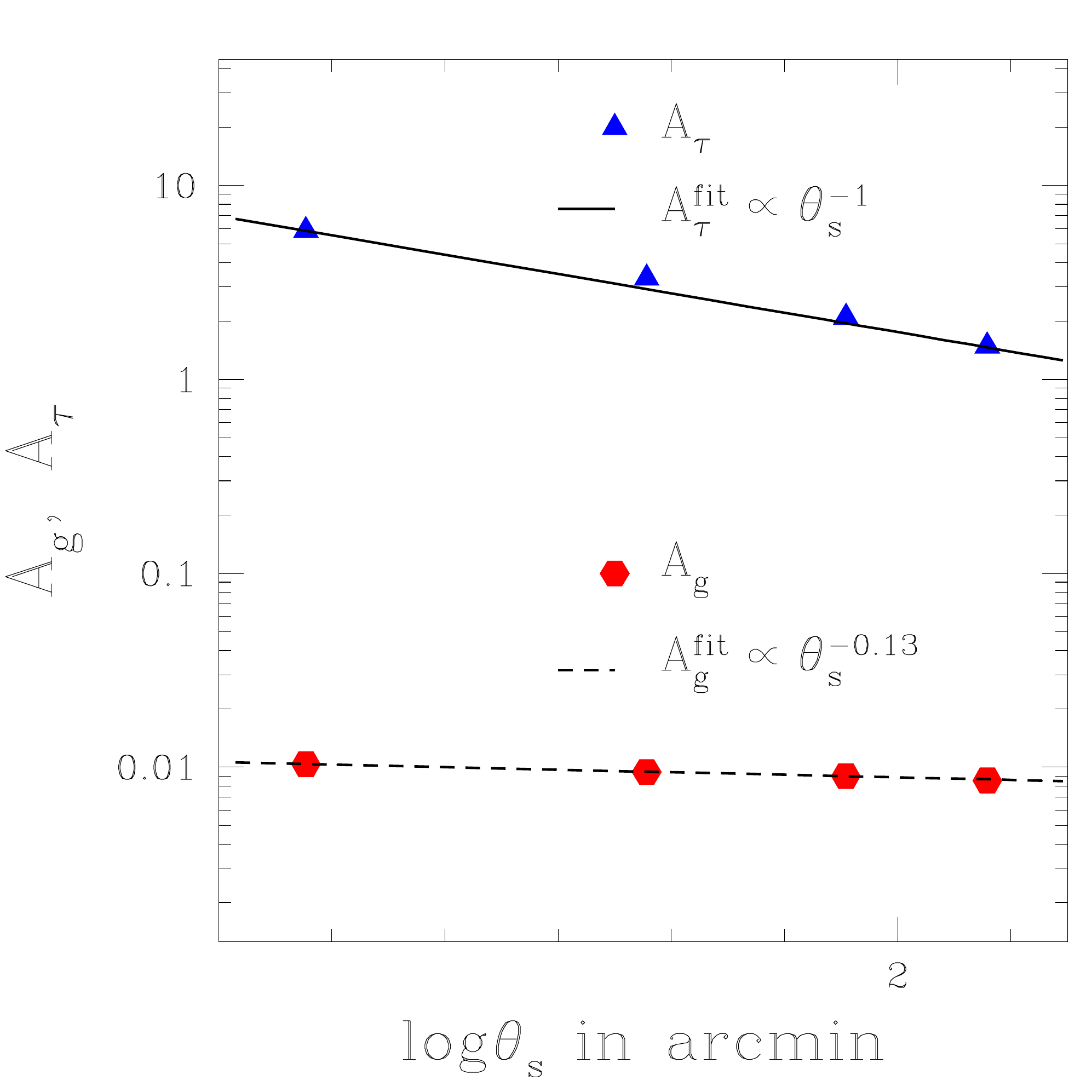}
\includegraphics[height=6cm,width=6cm]{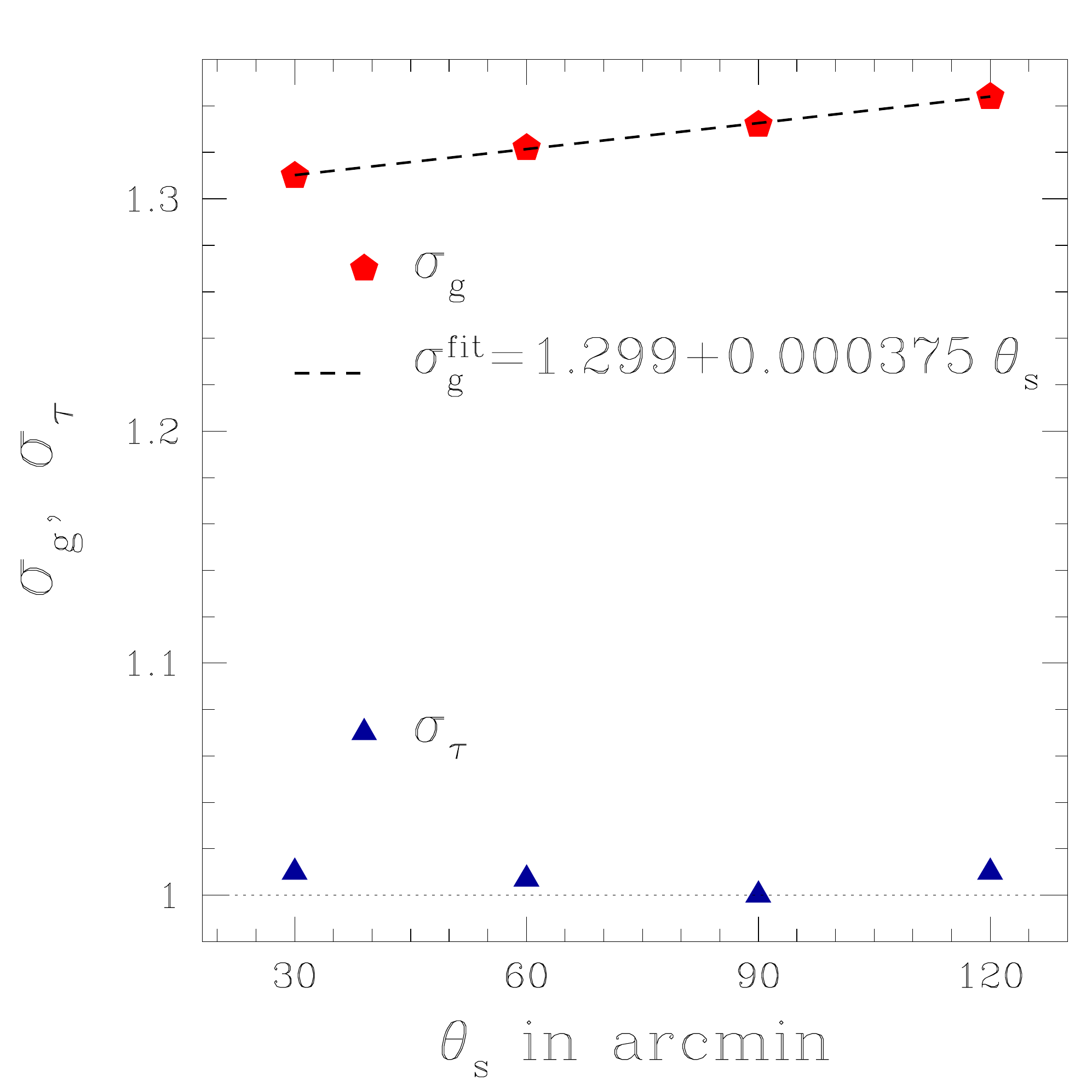}
\caption{{\em Top}: Scaling of amplitudes of $g$ and $\t$, denoted by $A_g$ and $A_{\t}$, respectively, with variation of the smoothing scale. {\em Bottom}: Scaling of $\sigma_g$ and $\s_{\tau}$ with $\theta_s$.}
\label{fig:2d_scaling}
\end{figure}

The parameters $A_g$, $\sigma_g$, $A_{\t}$ and $\s_{\t}$ are dependent on $s$. 
Fig.~\ref{fig:2d_scaling} shows how they  vary with $\theta_s$. 
As expected, $A_{\t}$ exhibits approximate power law behaviour. We find that $A_g$ also follows power law behaviour, which we express as,
\be A_g\propto \theta_s^{-\g}.\ee
The value of the exponent is obtained to be $\g=0.13$. Next, we find that $\s_{\t}$ is marginally larger than one at all smoothing scales, while $\s_g$ exhibits approximately linear increase with increase of $\theta_s$. 

{\em Functional form of $\la\a\ra$}:  We can now use the functional forms of   $\la g\ra $ and $\la \t\ra $ to determine the functional form of $\la \alpha \ra$. 
The ensemble mean of $\alpha (\nu)$ to first order in $\la g\ra /\la\t\ra$ is then given by
      \begin{equation}
       \la \alpha (\nu)\ra \simeq 1 - \frac{A_g}{A_{\t} }\,e^{\nu^2/2\Delta_g}, \label{eqn:alphamean}
      \end{equation}
where $\Delta_g\equiv 1/\left(\frac{1}{\s_{\t}^2}-\frac{1}{\sigma^2_g}\right)$. Since $\sigma_g>\s_{\t}$, we have $\Delta_g> 0$.      
This expression reproduces the shape of $ \la \alpha (\nu) \ra$ very well, as seen in the bottom panels of fig.~\ref{fig:2d_allnu} where this analytic fit is also plotted (red solid lines). Note that if we take $\s_{\t}$ to be exactly one, without accounting for the sampling effect, eqn.~\ref{eqn:alphamean} gives deviation from the numerical computation especially at higher thresholds. It is important to mention  that $g$ increases with decreasing $s$, due to increase of both $g_1$ and $g_2$. So, the manner in which $\alpha$ tends to one as $s\rightarrow 0$ is via $\tau$ increasing faster than $g$, and not via $g$ tending to zero. Therefore, the geometric meaning of isotropy for a random field is fundamentally different from that of  individual structures. 
     
\begin{figure}
 \centering\includegraphics[height=5.4cm,width=8.cm]{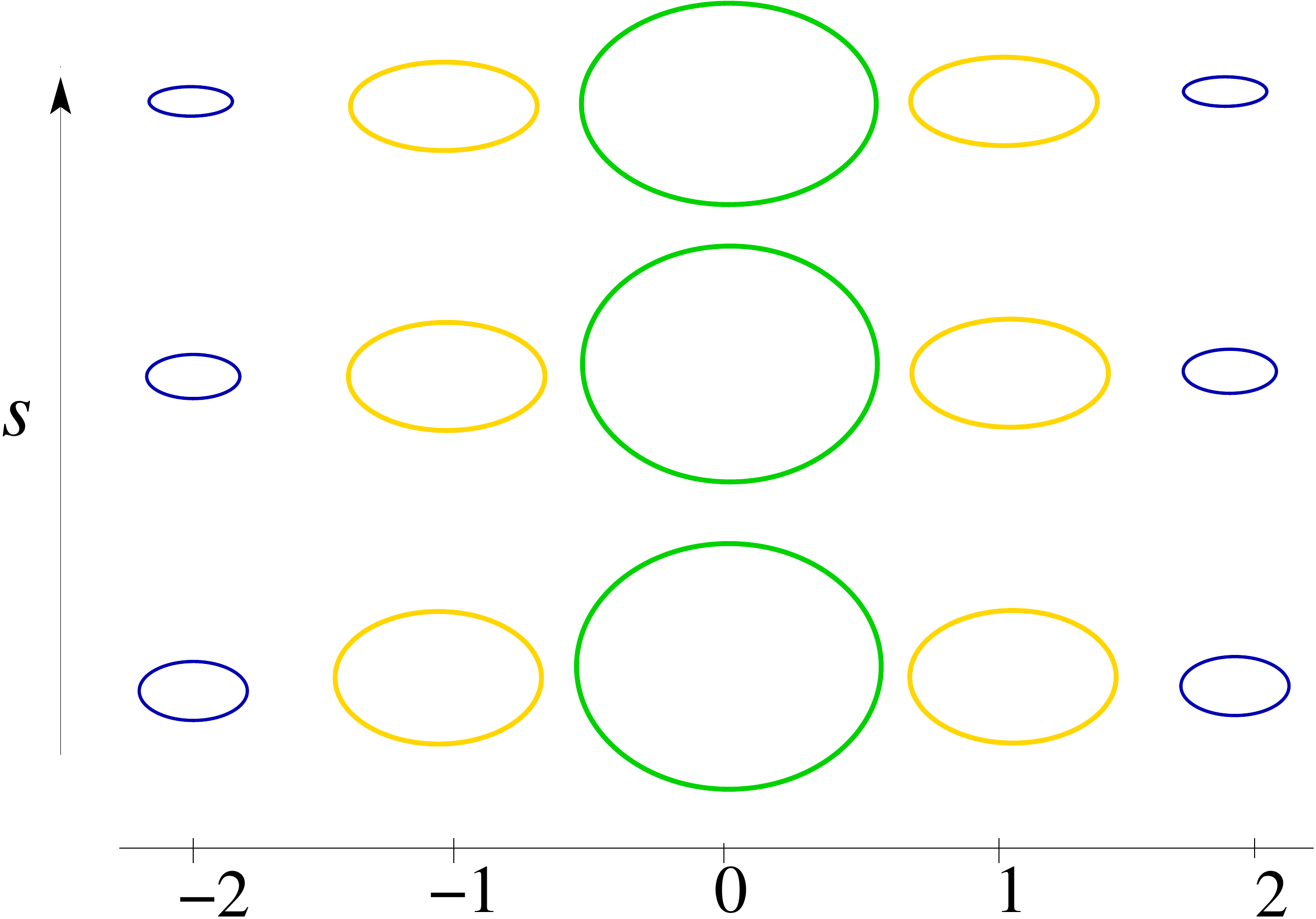}
  \vskip .2cm
\centering {\large{$\nu$}}
 \caption{Schematic representation of a Gaussian isotropic field for each resolution, $s$, as a series of ellipses at different thresholds. The vertical arrow shows the direction of increasing $s$.}
\label{fig:2disotropy}
\end{figure}

In fig.~\ref{fig:2disotropy} we show a schematic representation of a Gaussian isotropic field defined on a compact space and sampled at a finite number of pixels as a series of ellipses, with each ellipse corresponding to each $(s,\nu)$. The perimeters of the ellipses are Gaussian functions of $\nu$, and become more elliptic as $|\nu|\rightarrow\infty$, at each $s$. This representation captures the essence of statistical isotropy of smooth random fields at finite sampling. 

In appendix~\ref{sec:appen_pdfs} we discuss the probability density functions for $g_1$ and $\ g_2$,  and the resulting statistical nature of $g$ and $\varphi$. The results described in this section contain numerical errors introduced by the approximation of the delta function by the inverse of the threshold bin size~\cite{Lim:2011kd}. In appendix~\ref{sec:appen_err} we estimate the errors and show that they are small. 


\section{Ensemble expectation value of $\widetilde{\mathcal{W}}_1$ for Gaussian anisotropic fields}
\label{sec:sec5}

Let $f$ be Gaussian random field on the manifold $M$. Then $f_{;1}$ and $f_{;2}$ are also Gaussian fields. The joint PDF of $\mathbf{X} \equiv (f,f_{;1},f_{;2})$ is given by the Gaussian form
\bea
P(\mathbf{X})&=& \frac{1}{\sqrt{2\pi \,\texttt{Det}\mathbf\Sigma}}\, \exp\left(-\frac12 \mathbf{X}^T \mathbf{\Sigma}^{-1}  \mathbf{X}\right),
\eea
where $\mathbf{\Sigma}$  is the covariance matrix. 
Let $f$ be also given to be {\em anisotropic}. Then we can model the covariance matrix as 
\begin{equation}
  \mathbf{\Sigma}= \left(\begin{array}{ccc}
    \sigma_0^2 & 0 & 0 \\
     0& \sigma_{f_{;1}}^2 &  0  \\
     0&  0& \sigma_{f_{;2}}^2 
    \end{array}
    \right),
\end{equation}
where for $i=1,2$ we have used
\bea
\la f f_{;i}\ra = 0, \quad\la f_{;i}^2\ra \equiv \sigma_{f_{;i}}^2, \quad
\la f_{;1}f_{;2}\ra = 0.
\eea
For a general anisotropic field we have  $\sigma_{f_{;1}}^2\ne\sigma_{f_{;2}}^2$, while for isotropic case we can choose $\sigma_{f_{;i}}^2=\sigma_{f_{;2}}^2 =\s_1^2/2$,. We have taken the cross correlations 
$\la f f_{;i}\ra$ and $\la f_{;1}f_{;2}\ra$ to be zero even for anisotropic fields. This  corresponds to choosing appropriate coordinates for which these cross-correlations vanish. 

To keep the discussion general, we also consider  $\sigma_{f_{;1}}^2$ and $\sigma_{f_{;2}}^2$ to be dependent on the field threshold, $f=\nu\s_0$, keeping in mind that the finite sampling may introduce such threshold dependence, in addition to a difference between $\sigma_{f_{;1}}^2$ and $\sigma_{f_{;2}}^2$. we are assuming that the sampling effect does not  induce departure from Gaussian nature of the field, which need not be true. A discussion incorporating such a deviation is beyond the scope of this paper. 
Then $P_{\nu}(\mathbf{X})$ becomes 
\be
P_{\nu}(\mathbf{X})
= \frac{1}{\sqrt{2\pi\sigma_0^2 \sigma_{f_{;1}}^2\sigma_{f_{;2}}^2}} \exp\left\{-\frac12 \left( \frac{f^2}{\sigma_0^2} + \frac{f_{;1}^2}{\sigma_{f_1}^2} +  \frac{f_{;2}^2}{\sigma_{f_2}^2} \right) \right\},
\ee
where the index $\nu$ on $P$ is kept to remind us that it can be dependent on $\nu$ via $\mathbf{\Sigma}$.   
On the right hand side $\nu$ is not explicitly written. 

The ensemble expectation value of the $(i,j)$ element of ${\widetilde{\mathcal{W}}}_1$ at threshold $\nu$ is obtained to be
\begin{eqnarray}
 \langle {\widetilde{\mathcal W}}_1 \rangle &=&   
\frac{2\sqrt{2}}{\sigma_0}  \left(
 \begin{array}{ll}
     A_{1}F_{1} & 0\\
     0 & A_{2}F_{2}
   \end{array}
   \right) \,e^{-\frac{\nu^2}{2}},
  \label{eqn:Wex}
 \end{eqnarray}
 where 
  \begin{equation}
     A_{1} =  \frac{\sigma_{f_{;1}}^2\sigma_{f_{;2}}^2}{\left( 2\sigma_{f_{;2}}^2 - \sigma_{f_{;1}}^2\right)^{3/2} },\quad
     A_{2} =  \frac{\sigma_{f_{;1}}^2\sigma_{f_{;2}}^2}{\left( 2\sigma_{f_{;1}}^2 - \sigma_{f_{;2}}^2 \right)^{3/2}}. 
\end{equation}
The factor $F_i$, for $i=1,2$, is given by 
 \begin{equation}
   F_{i} =  \int_{0}^{\pi/2} \,{\rm d}y \frac{1}{\sqrt{D_i^2+ \tan^2 y}}   
 \frac{1}{\sqrt{\cos y}} \ \cos\left( \frac32 y \right),
 \end{equation}
 with
  \begin{eqnarray}
D_1=\frac{\sigma_{f_{;1}}^2}{2\sigma_{f_{;2}}^2-\sigma_{f_{;1}}^2},\quad  D_2=\frac{\sigma_{f_{;2}}^2}{2\sigma_{f_{;1}}^2-\sigma_{f_{;2}}^2}.  
\end{eqnarray}
 Eqn.~\ref{eqn:Wex} gives $ \la g_2\ra =0$. Let us denote $\bar g=|\la g_1\ra|$. Then we can express  $\la\tau\ra$ and $\bar g$ as
  \bea
\la  \tau(\nu)\ra &=&  \frac{\sqrt{2}}{\sigma_0} \bigg(A_1F_1 + A_2F_2\bigg) \,e^{-\frac{\nu^2}{2}}, \label{eqn:tauex}\\
  \bar g(\nu) &=& \frac{\sqrt{2}}{\sigma_0} \bigg(A_2F_2 - A_1F_1\bigg)  \,e^{-\frac{\nu^2}{2}}.
  \eea
Eq.~\ref{eqn:tauex} generalizes the well known expression for the second scalar MF, the contour length,  for Gaussian isotropic fields to   Gaussian but anisotropic fields.  

{\em Recovering the result for exact isotropy (limit $s\rightarrow 0$ if the anisotropy is due to sampling effect)}:  If  $ \sigma_{u_1}^2 = \sigma_{u_2}^2$, then eqn.~\ref{eqn:Wex} gives
 \begin{equation}
 \langle {\widetilde{\mathcal W}}_1 \rangle =   
   \frac{1}{8r_c} \,e^{-\nu^2/2} \times\,\mathbf{I} \times {\mathcal A},
  \label{eqn:Wex_iso}
 \end{equation}
 where $\mathbf{I}$ is the identity matrix, and $r_c$ is the correlation length of the field  given by $r_c=\sigma_0/\sigma_1$. This is the expression obtained in \cite{Chingangbam:2017uqv}.
 
From the ensemble expectation value of the rotated CMT, ${\widetilde{\mathcal{W}}}_1^R$, we recover the same expression for the expectation value of $\t$  as eqn.~\ref{eqn:tauex}. The ensemble expectation of $g$, on the other hand,  involves cross-correlations of $f_{;i}$ at different spatial locations. The full computation is beyond the scope of this paper. 


\section{Conclusion and discussion}
\label{sec:sec6}

In this paper we have addressed the question of statistical isotropy of smooth random fields in two dimensions from a  geometrical perspective. Before discussing random fields, we first carry out a detailed study of the geometry of single structure that is encoded in the CMT, building on our earlier work~\cite{Chingangbam:2017uqv}. We prove that the CMT  is proportional to the identity matrix for a smooth closed curve that has $m$-fold symmetry, with $m\ge 3$. The proportionality constant is just the perimeter of the curve. 
Next, we use the CMT to construct a mapping of an arbitrary closed curve to an ellipse that is unique up to translations of the centroid. Lastly, we show that the shape parameters that are defined using the CMT, and the filamentarity that is  defined using the scalar MFs (area and the perimeter), carry complementary shape information. Therefore, using a combination of both will maximize extraction of shape information in practical applications. However, the directionality information
that is inherent in the CMT cannot be obtained from the scalar MFs. 

Then we focus on excursion sets of simulated Gaussian isotropic CMB temperature maps, as examples of  smooth random fields on a compact space. In~\cite{Chingangbam:2017uqv} we had shown that we can associate a circle with each excursion set of the isotropic field such that the radius depends on the PDF of the field. Here we show that finite sampling, due to finite extent of the space and pixelization, introduces threshold dependent distortion of the circle associated with each excursion set, to an ellipse. Then we show that the parameter $g$, that encodes the sampling anisotropy, scales as a power law function of the sampling scale (equivalent to smoothing scale for our consideration here), with scaling index $\g=0.13$. We then use a semi-numerical approach  to obtain an analytic expression for the  shape parameter $\alpha$. We further extend the analytic derivation of the CMT of Gaussian isotropic fields to Gaussian anisotropic fields, and obtain expressions for the ensemble expectations of the perimeter and the anisotropy parameter $g$ (without sampling effect). 

Our results provide a deeper understanding of the effect of finite sampling on the statistical isotropy of random fields. 
The main underlying point is that a random field measured over a finite domain will always present some level of statistical anisotropy due to finite sampling, similarly to the violation of exact ergodicity within a finite volume. The contour Minkowski tensor  provides a measure of this anisotropy via the quantity $\alpha$. Exact isotropy is an idealized  state for which the Minkowski tensor is proportional to the identity matrix in every coordinate system, which results in $\alpha$ being one. That is, the symmetry  of the field is reflected in the structure of the tensor. This is true for both Gaussian and non-Gaussian fields. We have shown that both $\t$ and $g$ increase as the resolution of the sampling increases. Importantly, the manner in which the field approaches state of exact isotropy  in the limit of infinite resolution is by $\tau$ increasing faster than $g$, and not by $g$ approaching zero.

Actual anisotropic signals in the data can be distinguished from statistical anisotropy by comparing the eigenvalues of the tensor to its components in any given coordinate system. For an isotropic field defined on a finite domain, fluctuations of the off-diagonal components, and the inequality between diagonal elements, are drawn from the same distribution in any coordinate system. And their   magnitudes  are quantified by the difference between the  eigenvalues via $\alpha$. This knowledge allows us to infer the statistical significance of any detected anisotropic signal, beyond random fluctuations due to sampling.

Our analysis will be extended in the following directions. Our main results  are based on numerical calculations. The value of  $\gamma$ is expected to depend on cosmological parameters, and may depend on the curvature of $M$.  It would be interesting to understand $\gamma$ from first principles, and relate it to cosmological parameters and the geometry of $M$. 
The expression of $\a$ that we have obtained semi-numerically is specific to the Gaussian nature of the field. It will be interesting to  determine how it encodes the nature of the field for other types of fields. 
The extension of the analysis carried out in this paper to random fields on three dimensional space will be the subject of our upcoming paper.


\section*{Acknowledgment}{We acknowledge the use of the \texttt{NOVA} HPC cluster at the Indian Institute of Astrophysics. Some of the results in this paper have been obtained by using the \texttt{CAMB}~\cite{Lewis:2000ah,cambsite} and \texttt{HEALPIX}~\cite{Gorski:2005,Healpix}  packages. PC~would like to thank Pankaj Sharan for the many lectures and discussions on  differential geometry. PC~ would also like to thank IISER Mohali for a visit during which a part of this work was carried out.  The work of PC~is supported by the Science and Engineering Research Board of the Department of Science and Technology, India, under the \texttt{MATRICS} scheme, bearing project reference no \texttt{MTR/2018/000896}.  SA is supported by an appointment to the JRG Program at the APCTP through the Science and Technology Promotion Fund and Lottery Fund of the Korean Government, and was also supported by the Korean Local Governments in Gyeongsangbuk-do Province and Pohang City. }

\appendix
\section{Probability density functions of $g_1$, $g_2$, $g$ and $\varphi$}
\label{sec:appen_pdfs}

If the variables $g_1$ and $g_2$ have Gaussian probability density functions (PDFs), then the PDF of $g$ will be Rayleigh form and that of $\varphi$ will be uniform. Here, we compute the PDFs numerically and show that $g_1$ and $g_2$ have approximately Gaussian PDFs.

The integrand for $g_1$ is the random field $q_1(\vx)\equiv (u_{;1}^2 -u_{;2}^2)/|\nabla u|$, while that of $g_2$ is $q_2(\vx)\equiv u_{;1}u_{;2}/|\nabla u|$. We can express the numerator of  $q_2$ as a linear combination 
\be
u_{;1}u_{;2} = \frac14 \left\{(u_{;1}+u_{;2})^2 - (u_{;1}-u_{;2})^2\right\}.
\ee
If $u_{;1},\,u_{;2}$ at each $\vx$ are  Gaussian variables with the same values of the mean and variance, then their sum and difference are also Gaussian variables.  Therefore, $q_1$ and $q_2$, and consequently $g_1$ and $g_2$, must have identical PDFs. 

\begin{figure}
\includegraphics[height=2.5cm,width=2.7cm]{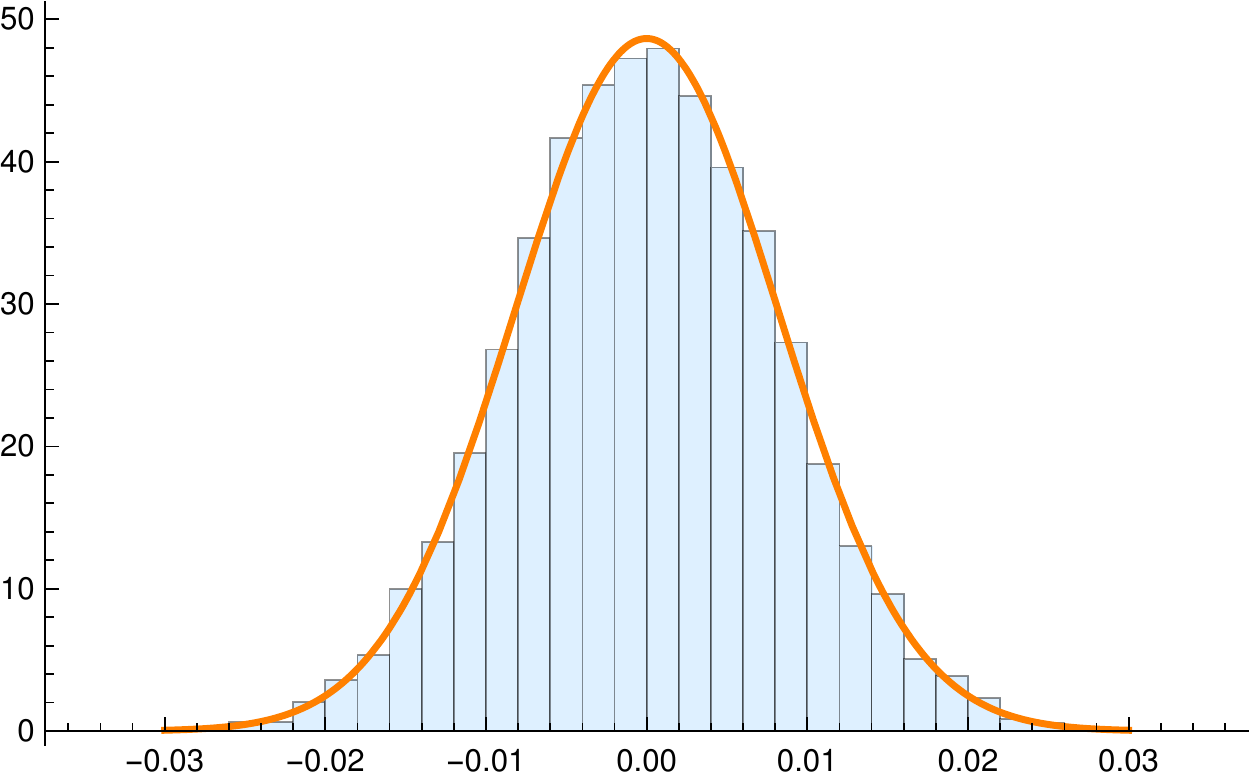}
 \includegraphics[height=2.5cm,width=2.7cm]{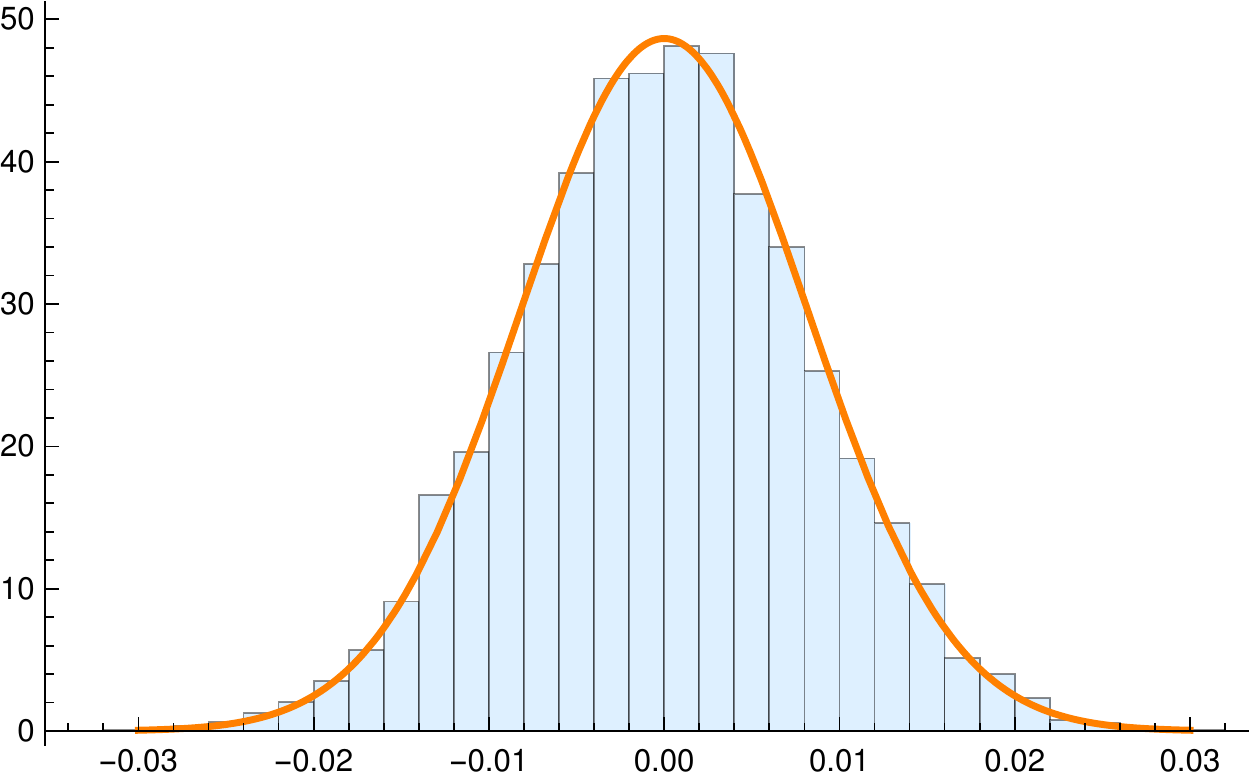}
 \includegraphics[height=2.5cm,width=2.7cm]{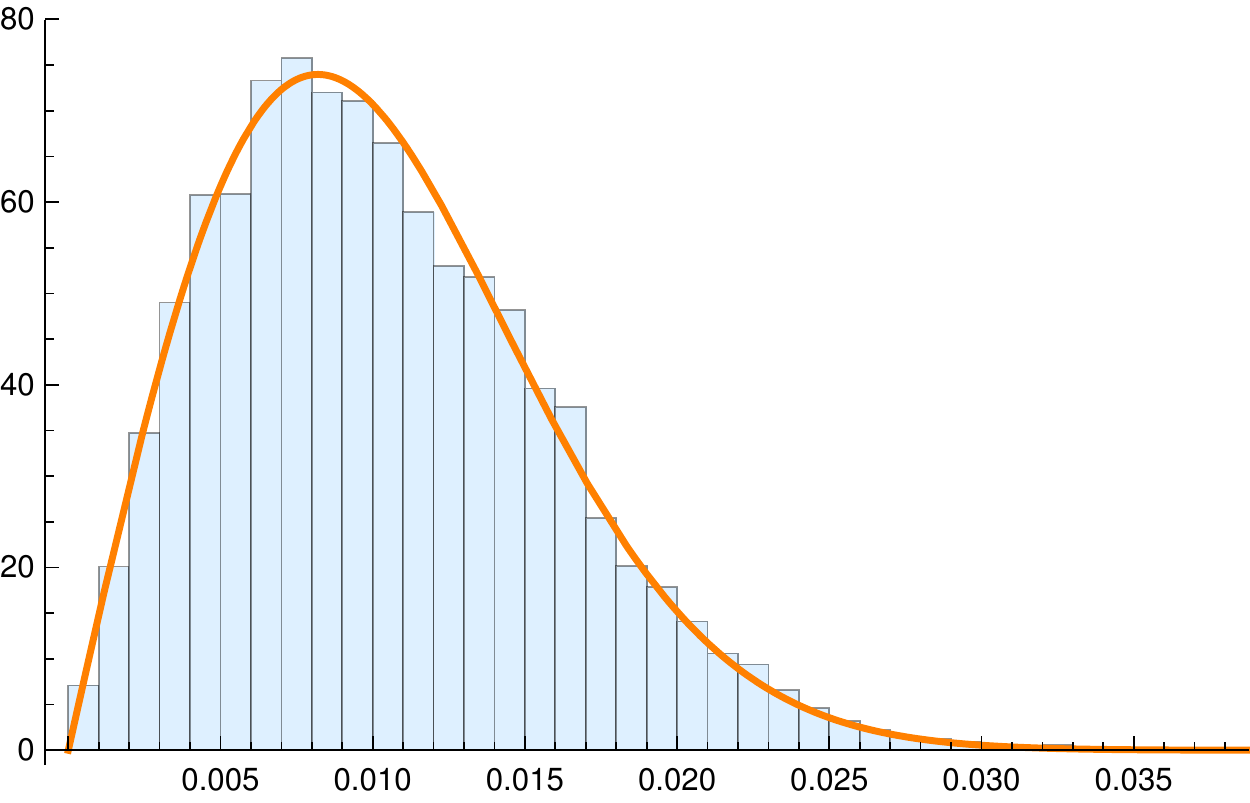}\\
 \centering{$g_1$ \hskip 2.3cm $g_2$ \hskip 2.3cm $g$}\\ \vskip .3cm
   \caption{PDFs of $g_1$, $g_2$ and $g$ obtained from  $10^4$ CMB temperature simulations. The smoothing scale is $\theta_s=60'$ and threshold value is $\nu=0$.  $g_1$, $g_2$ are well fit by identical Gaussian PDFs (orange solid lines) with zero mean, while $g$ is fit by the corresponding Rayleigh distribution (orange solid line).}
\label{fig:pdfs}
\end{figure}

Fig.~\ref{fig:pdfs} shows the PDFs  (yellow bars) of $g_1$, $g_2$ and $g$    
obtained using $10^4$ simulations of Gaussian isotropic CMB temperature. The smoothing scale is $\theta_s=60'$, and the threshold value is $\nu=0$. 
We find that $g_1$, $g_2$ are approximated very well by identical Gaussian PDFs (orange solid lines), with standard deviation value  $0.0074$. 
$g$ is fit by the corresponding Rayleigh distribution  (orange solid line).  The variance of $g_1$ and $g_2$ will vary with the threshold value and with the smoothing scale. 

The numerically computed PDFs of $g_1$ and $g_2$  will actually deviate from the exact Gaussian forms due to finite sampling. This will result in $\varphi$ deviating from uniform distribution, which will then leads to the value of $\alpha$ deviating from one. Analytic expressions for the PDFs of $g_1$, $g_2$ and the other related statistics  will be discussed in a separate work.

\section{Numerical error due to threshold binning on approximation of the $\d$ function}
\label{sec:appen_err}

The threshold binning introduces numerical error in the calculation of $\widetilde\W_1$ due to the approximation of the delta function~\cite{Lim:2011kd}. In~\cite{Goyal:2019vkq} we had argued why this error is expected to be small when computing $\alpha$. Here we make the argument quantitative and estimate the error as follows.

Let $h$ represent the numerically calculated value of either of the four quantities - $\la g_1\ra$, $\la g_2\ra$, $\la g\ra$ or $\la \t\ra$. Then writing $h$ as the sum of `true'  and `error' components, we get 
\be
\Delta h^{\rm err}(\nu)= h(\nu)-h^{\rm true}(\nu).
\label{eqn:A1}
\ee
The numerically calculated $h$ can be expressed as
\be
h(\nu)=\frac{1}{\D\nu} \int_{\nu-\Delta\nu/2}^{\nu+\Delta\nu/2} {\rm d}\nu' \, h^{\rm true}(\nu').
\ee
Let $h^{\rm true}(\nu)=Ae^{-\nu^2/2\s^2}$. Then inserting $h^{\rm true}(\nu)$ and $h(\nu)$ in eqn.~\ref{eqn:A1} we get,
\bea
\Delta h^{\rm err}(\nu) &=& \sqrt{\frac{\pi}{2}} \frac{\s A}{\D\nu} \bigg\{ {\rm erf}\left( \frac{\nu+\Delta\nu/2}{\sqrt{2}\s}\right) \nn\\
&& - {\rm erf}\left(\frac{\nu-\Delta\nu/2}{\sqrt{2}\s}\right) \bigg\}   -\,h^{\rm true}(\nu).
\eea
This gives the fractional error to be
\bea
\frac{\Delta h^{\rm err}}{h^{\rm true}}(\nu) &=& \sqrt{\frac{\pi}{2}} \frac{\s}{\D\nu} e^{\nu^2/2\s^2}\bigg\{ {\rm erf}\left( \frac{\nu+\Delta\nu/2}{\sqrt{2}\s}\right) - \nn\\
&& {\rm erf}\left(\frac{\nu-\Delta\nu/2}{\sqrt{2}\s}\right) \bigg\} -\,1.
\label{eqn:frac_error}
\eea
We can see that the right hand side of eqn.~\ref{eqn:frac_error} depends only on the threshold bin size and $\s$, and is independent of the amplitude of $h$. This, combined with the fact that the values of $\s_g$ and $\s_{\t}$ are comparable and of order one, implies that the errors of all four quantities - $\la g_1\ra$,  $\la g_2\ra$, $\la g\ra$ and $\la \t\ra$ - will be comparable.

\begin{figure}
 \centering\includegraphics[height=7cm,width=8cm]{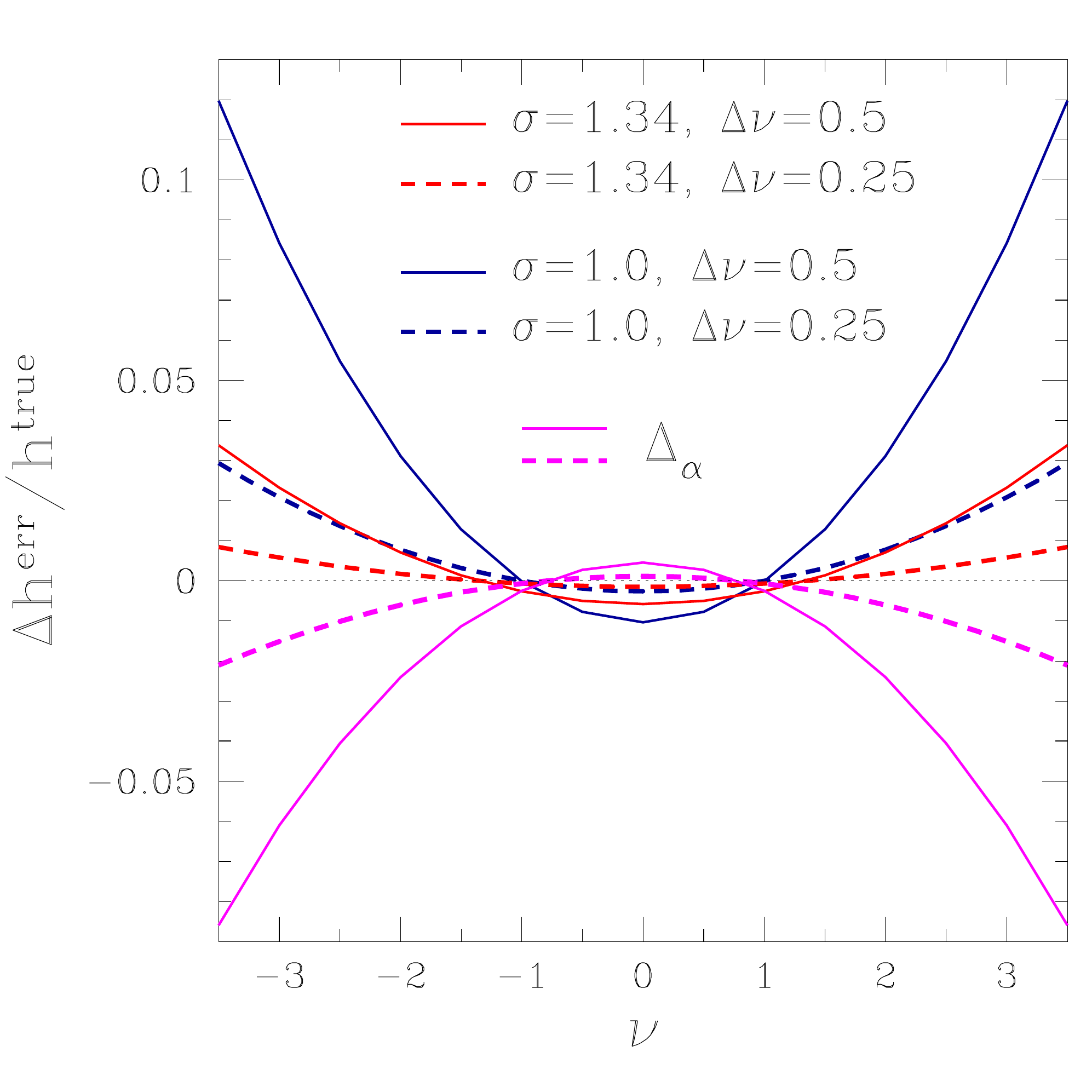}
   \caption{Plots of the fractional error given by eqn.~\ref{eqn:frac_error} for two different values of $\s$ and $\D\nu=0.25$ and 0.5. The magenta lines  represent $\Delta_{\a}$ which is the corresponding difference between the blue and red lines,  defined by  eqn.~\ref{eqn:Dalpha}.}
\label{fig:error}
\end{figure}

Fig.~\ref{fig:error} shows $\Delta h^{\rm err}/h^{\rm true}$ versus $\nu$, for two values $\Delta\nu=0.25,0.5$,  and for $\s=1$ and 1.34 (which are roughly the values relevant for section~\ref{sec:sec4B}). We see that the error is smaller for larger value of $\s$. So the fractional error for calculating $\t$ is larger than for $g$. The fractional error also decreases as we decrease $\Delta\nu$.  

The error for $\la g\ra/\la \t\ra$ can now be estimated as
\be
\frac{\la g\ra}{\la \t\ra} \simeq \frac{\la g\ra^{\rm true}}{\la \t\ra^{\rm true}} \bigg( 1+\frac{\Delta g^{\rm err}}{g^{\rm true}} - \frac{\Delta \t^{\rm err}}{\t^{\rm true}}  \bigg).
\ee
Let us denote the sum of the last two terms as 
\be
\Delta_{\a}=\frac{\Delta g^{\rm err}}{g^{\rm true}} - \frac{\Delta \t^{\rm err}}{\t^{\rm true}}.
\label{eqn:Dalpha}
\ee
$\Delta_{\a}$ is shown by the magenta lines, solid for $\D\nu=0.5$ and dashed for 0.25, in fig.~\ref{fig:error}. We see that it is sub-percent towards smaller threshold and for smaller $\Delta\nu$.

There can be small residual error, in addition to $\D_{\a}$,  due to inaccuracy in the determination of $\s$ and the difference between the fractional errors of $g$ and $\t$ which we ignore here.


\end{document}